\documentclass[preprint,amsmath,amssymb,aps,a4]{revtex4-2}
\usepackage{graphicx}
\usepackage{bm}
\usepackage{txfonts}
\usepackage{hyperref}
\date{\today}
\newcommand{\be}{\begin{eqnarray}}
\newcommand{\ee}{\end{eqnarray}}

\newcommand{\bfp}{{\bf p}_{\perp}}

\usepackage{xcolor}
\begin{document}
\title{Twist-4 T-even proton TMDs in the light-front quark-diquark model}
\author{Shubham Sharma}
\email{s.sharma.hep@gmail.com}
\affiliation{Department of Physics, Dr. B. R. Ambedkar National Institute of Technology, Jalandhar 144027, India}
\author{Harleen Dahiya}
\email{dahiyah@nitj.ac.in}
\affiliation{Department of Physics, Dr. B. R. Ambedkar National Institute of Technology, Jalandhar 144027, India}

\date{\today}%
\begin{abstract}
We have dealt with the twist-4  T-even transverse momentum dependent parton distributions (TMDs) for the case of proton in the light-front quark-diquark model (LFQDM). By decoding the unintegrated quark-quark correlator for the semi-inclusive deep inelastic scattering (SIDIS), we have specifically obtained the overlap form for the unpolarized \bigg($f_{3}^{\nu}(x, {\bf p_\perp^2})$\bigg), longitudinally polarized \bigg($ g_{3L}^{\nu}(x, {\bf p_\perp^2}),~h_{3L}^{\perp\nu}(x, {\bf p_\perp^2})$\bigg) and transversely polarized \bigg( ${g}^{\nu  }_{3T}(x, {\bf p_\perp^2}),~{h}^{\nu  }_{3T}(x, {\bf p_\perp^2})$ and ${h}^{\nu\perp}_{3T}(x, {\bf p_\perp^2})$\bigg) proton TMDs. We have provided the explicit expressions for both the cases of the diquark being a scalar or a vector. Average transverse momenta and the average square transverse momenta for the TMDs have been calculated and the results have been tabulated with corresponding leading twist TMDs. In addition, the value of average transverse momentum and average square transverse momentum for TMD ${f}^{\nu  }_3(x, {\bf p_\perp^2})$ has been compared with the available light-front constituent quark model (LFCQM) results.
From TMDs, we have also obtained and discussed the transverse momentum dependent parton distribution functions (TMDPDFs). The model relations of the twist-4 T-even TMDs with the available leading twist T-even TMDs have also been obtained.
\par
 \vspace{0.1cm}
\end{abstract}
%
\maketitle
%
%
\section{Introduction\label{secintro}}
The fact that transverse momentum dependent parton distributions (TMDs) are a generalization \cite{Collins:2003fm,Collins:2007ph,Collins:1999dz,Hautmann:2007uw} of parton distribution functions (PDFs) is one of the main factors contributing to their appeal to particle physicists. TMDs appear to have the potential to improve our understanding of the structure of the nucleon beyond what we now know from PDFs, namely, the distribution of partons in the longitudinal momentum space. TMDs encrypt the information concerning spin-orbit correlations, the angular momentum of the nucleon, and the three-dimensional (3-D) structure \cite{Collins:1981uk,Ji:2004wu,Collins:2004nx,Cahn:1978se,Konig:1982uk,Chiappetta:1986yg,Collins:1984kg,Sivers:1989cc,Efremov:1992pe,Collins:1992kk,Collins:1993kq,Kotzinian:1994dv,Mulders:1995dh,Boer:1997nt,Boer:1997mf,Boer:1999mm,Bacchetta:1999kz,Brodsky:2002cx,Collins:2002kn,Belitsky:2002sm,Burkardt:2002ks,Pobylitsa:2003ty,Goeke:2005hb,Bacchetta:2006tn,Cherednikov:2007tw,Brodsky:2006hj,Avakian:2007xa,Miller:2007ae,Arnold:2008kf,Brodsky:2010vs,lattice-TMD}.
\par 
Transverse momentum dependent fragmentation functions and
TMDs are useful to go into the detailed explanation of leading twist observables in the deep inelastic scattering (DIS) experiments
\cite{Collins:1981uk,Ji:2004wu,Collins:2004nx}
on which we have accessible data from hadron production in $e^+e^-$ annihilation \cite{Abe:2005zx,Ogawa:2006bm,Seidl:2008xc,Vossen:2009xz}, Drell-Yan process \cite{Falciano:1986wk,Conway:1989fs,Zhu:2006gx} and semi-inclusive deep inelastic scattering (SIDIS) processes \cite{Arneodo:1986cf,Airapetian:1999tv,Avakian:2003pk,Airapetian:2004tw,Alexakhin:2005iw,Gregor:2005qv,Ageev:2006da,Airapetian:2005jc,Kotzinian:2007uv,Diefenthaler:2005gx,Airapetian:2008sk,Osipenko:2008rv,Giordano:2009hi,Gohn:2009,Airapetian:2009jy}.
Magnificent progress has been done in the higher order QCD calculations \cite{Gehrmann:2014yya,Echevarria:2015byo,Echevarria:2016scs,Li:2016ctv,Vladimirov:2016dll,Gutierrez-Reyes:2017glx,Gutierrez-Reyes:2018iod,Luo:2019hmp,Luo:2019szz,Ebert:2020yqt} and phenomenological studies
\cite{Efremov:2004tp,Anselmino:2005nn,Vogelsang:2005cs,Collins:2005ie,Collins:2005rq,Anselmino:2007fs,Anselmino:2013vqa,Signori:2013mda,Anselmino:2013lza,Kang:2014zza,Kang:2015msa,Kang:2017btw,Cammarota:2020qcw,Lefky:2014eia}.
Besides this, different aspects of TMD physics have been reviewed in detail \cite{Collins:2003fm,DAlesio:2007bjf,Barone:2010zz,Aidala:2012mv,Avakian:2019drf,Anselmino:2020vlp}. The transverse structure of the nucleon has been explored with the help of TMDs in the SIDIS processes and it is one of the key motivations behind the Electron-Ion Collider (EIC) \cite{Accardi:2012qut,EIC2103.05419,EIC2203.13199,EIC2203.13923}. \par
Being non-perturbative by nature, TMDs are calculated in various quantum chromodynamics (QCD) inspired models. Distributions of hadrons (TMDs, PDFs, form factors distributions etc.) have been studied in quark-target
\cite{Kundu:2001pk,Meissner:2007rx,Mukherjee:2009uy,Mukherjee:2010iw,Xu:2019xhk}, holographic \cite{Maji:2017wwd,Lyubovitskij:2020otz}, Nambu--Jona-Lasinio \cite{Matevosyan:2011vj}, Valon \cite{Yazdi:2014zaa}, light-front constituent quark (LFCQM) \cite{Pasquini:2008ax,Pasquini:2010af,Lorce:2011dv,Boffi:2009sh,Pasquini:2011tk,Lorce:2014hxa,Kofler:2017uzq,Pasquini:2018oyz}, quark-diquark \cite{Jakob:1997wg,Gamberg:2007wm,Cloet:2007em,Bacchetta:2008af,
She:2009jq,Lu:2012gu,Maji:2015vsa,Maji:2016yqo,Maji:2017bcz}, covariant parton model (CPM) \cite{Bastami:2020rxn}, chiral quark soliton \cite{Diakonov:1996sr,Diakonov:1997vc,Gamberg:1998vg,Pobylitsa:1998tk,Goeke:2000wv,Wakamatsu:2000fd,Schweitzer:2001sr,Schweitzer:2003uy,Wakamatsu:2003uu,Ohnishi:2003mf,Cebulla:2007ej,Wakamatsu:2009fn,
Schweitzer:2012hh} and bag models \cite{Jaffe:1991ra,Yuan:2003wk,Courtoy:2008vi,Avakian:2008dz, Courtoy:2008dn,Avakian:2010br}. In spectator model framework, gluon distributions like T-even, T-odd gluon Sivers and linearity TMDs  have been calculated by considering a spin-$1/2$ triquark \cite{BCR2107.13446,BCR2111.01686,BCR2111.03567,BCR2201.10508,BCRT2005.02288,C2101.04630,C2202.04207}. In addition to this, model-independent lattice QCD computations have been done in some cases \cite{lattice-TMD,Musch:2010ka,Musch:2011er,
Chen:2016utp,Alexandrou:2016jqi,Yoon:2017qzo,Orginos:2017kos,Joo:2019jct}. \par
Light-front AdS/QCD for a two-body bound state has predicted a multitude of exciting nucleon features \cite{BT,PRD83,PRD89,PRD91,EPJC77}. It is compatible with Drell-Yan-West relation \cite{PRD89,DY70,West70} and the quark counting rule \cite{Maji:2016yqo,PRD89}. Considering light-front AdS/QCD, first calculation of GPDs have been done in Ref. \cite{PRD83}. Additionally. from a unified point of view, calculations of TMDs, Wigner and Husimi  distributions have also been performed \cite{EPJC77}. In the light-front quark-diquark model (LFQDM), proton is described as a composite of an active quark and a diquark spectator of definite mass \cite{Chakrabarti:2019wjx}. The light-front wave functions (LFWFs) constructed from the AdS/QCD predictions involve the contributions from the scalar ($S=0$) and axial vector ($S=1$) diquarks and it has a $SU(4)$ spin-flavor structure \cite{Maji:2016yqo}. It is suitable for the PDF evolution from $\mu^2=0.09 ~\mathrm{GeV}^2$ to any arbitrary scale (may be extremely high i.e., upto $\mu^2=10^4~ ~\mathrm{GeV}^2$ ), therefore, we can calculate the distributions at any random scale. The transversity and helicity PDFs have been analyzed in this model. It has been shown that Soffer bound is in accordance with the available data. The experimental values of axial and tensor charges have been reproduced in this model. In this model, generalized parton distributions (GPDs) have been investigated for the quarks in a proton in both momentum and position spaces \cite{Mondal:2015uha}. The outcomes have been compared to the soft-wall AdS/QCD model for proton GPDs with zero skewness. GPDs for nonzero skewness have also been computed. GPDs show a diffraction pattern in longitudinal position space, which has also been shown by other models \cite{Mondal:2015uha}. Comparative study of the nucleon charge and anomalous magnetization density in the transverse plane have been done \cite{Mondal:2015uha}. Flavor decompositions of the transverse densities and form factors have also been performed \cite{Mondal:2015uha}.

Leading twist T-odd quark TMDs of the proton have also been obtained, specifically the Sivers function $f_{1T}^{\perp q}(x, {\bf p_\perp^2})$ and the Boer-Mulders function $h_{1}^{\perp q}(x, {\bf p_\perp^2})$ \cite{Gurjar:2022rcl}. The generalized Sivers and Boer-Mulders shifts have also been compared to the known lattice QCD simulations and it has been proved that the SIDIS spin asymmetries associated with these T-odd TMDs are consistent with HERMES and COMPASS findings \cite{Gurjar:2022rcl}.
Leading twist T-even TMDs have been discussed in this model including both scalar and vector diquark \cite{Maji:2017bcz}. Inequalities, common to diquark models, are found to be satisfied in this model as well. Soffer bound is found to be satisfied by the transversity TMD. Even though, $x$-$p_\perp$ factorization is not observable in this model, contrary to the other phenomenological models for the TMD $f_1^\nu(x,{p_\perp}^2)$, numerical analysis of TMDs are in sync with the phenomenological ansatz. Transverse shape of the  proton has been successively presented in this model. For leading twist TMDs, this model has shown the relation of quark densities with the first moments in $x$ of TMDs $f_1^\nu(x,p_\perp^2)$ and $g_{1T}^\nu(x,p_\perp^2)$ for different polarization of quarks and the parent proton. PDFs are obtained by successively $p_\perp$-integration, $f_1^\nu(x,p_\perp^2),~ h_1^\nu(x,p_\perp^2)$ and $g_{1L}^\nu(x,p_\perp)$ give the PDFs $f_1^\nu(x),~h_1^\nu(x)$ and $g_1^\nu(x)$ while there is no such collinear interpretations for the other TMDs. When integrated TMDs are processed in this model through Dokshitzer-Gribov-Lipatov-Altarelli-Parisi (DGLAP) evolution at the high scales, some specific ratios such as $g_{1T}^\nu(x)/h_{1L}^\nu(x)$ and $h_{1T}^\nu(x)/f_1^\nu(x)$ do not depend on the evolution scale $\mu$. It would be important to mention that this link with collinear factorization is a model-dependent property which holds only at tree level and at an initial energy scale. After that TMD densities and their collinear counterparts decouple due to different evolution equations, Collins-Soper-Sterman (CSS) versus DGLAP. The relationships between TMDs and GPDs in a LFQDM have been studied and many of the relationships which we have obtained seems to have a similar structure in several models \cite{Gurjar:2021dyv}. The relationship between the Sivers function and the GPD $E_q$ can be derived in terms of a lensing function in this model \cite{Gurjar:2021dyv}. The orbital angular momentum of quarks is computed and compared to the results of other similar models \cite{Gurjar:2021dyv}. 
Gravitational form factors (GFFs) have been obtained in this model and their consequences on the mechanical property description, have been studied, i.e. the mechanical radius, shear forces inside the proton and the distributions of pressures \cite{Chakrabarti:2020kdc}. The GFFs, $A(Q^2)$ and $B(Q^2)$ obtained are consistent with the lattice QCD, whereas the qualitative behaviour of the D-term form factor is consistent with the extracted data from the deeply virtual Compton scattering (DVCS) experiments at JLab, the lattice QCD, and the predictions of various phenomenological models \cite{Chakrabarti:2020kdc}. The distributions of pressure and shear force are also compatible with the outcomes of other models \cite{Chakrabarti:2020kdc}.\par
Higher twist (subleading twist and twist-4) TMDs have been studied for hadrons \cite{Avakian:2010br,Jakob:1997wg,Lorce:2014hxa,
Pasquini:2018oyz,Kundu:2001pk,Mukherjee:2010iw}. Twist-4 distributions have been the topic of interest \cite{Lorce:2014hxa,PhysRevLett.67.552,SIGNAL1997415,PhysRevD.95.074017,ELLIS19821,ELLIS198329,QIU1991105,QIU1991137,PhysRevD.83.054010} specifically the twist-4 T-even unpolarized TMD  $f_{3}^{\nu}(x, {\bf p_\perp^2})$ \cite{Lorce:2014hxa}. Even though the twist-4 distributions have been discussed in Ref. \cite{PhysRevD.95.074017} and in the spectator model \cite{liu21}, there is still a need to study twist-4 TMDs in detail i.e., representation of their variation with TMD variables, average transverse momentum values, relations with leading twist variables and their transverse momentum dependent parton distribution functions (TMDPDFs).
\par
The purpose of the present work is to study the twist-4 T-even TMDs for the case of  proton in the framework of LFQDM. For SIDIS, we have decrypted the unintegrated quark-quark correlator and obtained the overlap form of the TMDs for desired polarization of proton. Their explicit expressions are provided for both the cases of diquark being a scalar or vector. We have discussed the 2-D and 3-D variation of these TMDs for both the $u$ and $d$ quarks, with longitudinal momentum fraction $x$ and transverse momentum ${\bf p_\perp^2}$. We have plotted the TMDPDFs which have been obtained by integrating the TMDs over the transverse momentum of quark ${\bf p_\perp}$. We have also debated over the comparison of TMDPDFs plots with the 2-D plots of TMDs verses $x$. Since this twist is less explored, in order to achieve an enhanced absorption of the nature of derived TMDs, we have expressed our result of twist-4 T-even TMDs in the form of available leading twist T-even TMDs. To be more precise, the unpolarized twist-4 T-even TMD $f_{3}^{\nu}(x, {\bf p_\perp^2})$ is expressed in the form of unpolarized leading twist T-even TMD ${f}^{\nu  }_1(x, {\bf p_\perp^2})$, twist-4 T-even longitudinally polarized TMDs $\bigg(g_{3L}^{\nu}(x, {\bf p_\perp^2})$ and $ h_{3L}^{\perp\nu}(x, {\bf p_\perp^2})\bigg)$ in the form of leading twist T-even longitudinally polarized TMDs  $\bigg({h}^{\nu\perp}_{1L}(x, {\bf p_\perp^2})$ and ${g}^{\nu  }_{1L}(x, {\bf p_\perp^2})\bigg)$ and the transversely polarized twist-4 T-even TMD $\bigg({h}^{\nu  }_{3T}(x, {\bf p_\perp^2}),~{g}^{\nu  }_{3T}(x, {\bf p_\perp^2})$ and ${h}^{\nu\perp}_{3T}(x, {\bf p_\perp^2})\bigg)$ are expressed in the form of leading twist T-even transversely polarized TMDs  $\bigg({g}^{\nu  }_{1T}(x, {\bf p_\perp^2}),~{h}^{\nu\perp}_{1T}(x, {\bf p_\perp^2}) $ and ${h}^{\nu}_{1}(x, {\bf p_\perp^2})\bigg) $ within the same model. We have also tabulated the results of average transverse momenta and average transverse momenta square for our twist-4 T-even TMDs and compared it with their corresponding leading twist TMDs in LFQDM. Further, we have compared our result for TMD ${f}^{\nu  }_3(x, {\bf p_\perp^2})$ with LFCQM results.
\par
The work is arranged as follows: Essential details of the quark-diquark model have been discussed in Sec.\ref{secmodel}. Input parameters of the model have been given in Sec.\ref{secinputp}. In Sec.\ref{sectmdsp}, the twist-4 quark TMDs projections have been shown in the form of quark-quark correlator. We have presented our results of the twist-4 T-even TMDs in the overlap form of LFWFs along with their explicit expressions and their relation with leading twist T-even TMDs in Sec.\ref{secresults}. Interpretation of TMDs with the help of 2D, 3D-Plots and its TMDPDFs has been done in Sec.\ref{secdiscussion}. Average transverse momenta and average transverse momenta square for our twist-4 T-even TMDs and its comparison with prior results has also been done in this section. The results have been concluded in Sec.\ref{seccon}.
\section{Light-Front Quark-Diquark Model \label{secmodel}}
In this section, we have presented the essentials of the LFQDM \cite{Maji:2016yqo} where a  proton is described as a composite of an active quark and a diquark spectator of definite mass \cite{Chakrabarti:2019wjx}. 
The spin-flavor $SU(4)$ structure of the proton can be expressed as an aggregate of isoscalar-scalar diquark singlet $|u~ S^0\rangle$, isoscalar-vector diquark $|u~ A^0\rangle$ and isovector-vector diquark $|d~ A^1\rangle$ states as \cite{Jakob:1997wg,Bacchetta:2008af}
\begin{equation}
|P; \pm\rangle = C_S|u~ S^0\rangle^\pm + C_V|u~ A^0\rangle^\pm + C_{VV}|d~ A^1\rangle^\pm. \label{PS_state}
\end{equation}
Here, the scalar and vector diquark has been represented by $S$ and $A$ respectively. Isospin of the diquark has been shown by the superscripts on them. The light-cone convention $z^\pm=z^0 \pm z^3$ has been used and the frame is picked such that the proton's transverse momentum disappear i.e., $P \equiv \big(P^+,\frac{M^2}{P^+},\textbf{0}_\perp\big)$. The momentum of the thrashed quark ($p$) and diquark ($P_X$) are
\be
p &&\equiv \bigg(xP^+, \frac{p^2+|\bfp|^2}{xP^+},\bfp \bigg),\\
P_X &&\equiv \bigg((1-x)P^+,P^-_X,-\bfp\bigg).
\ee
The representation of longitudinal momentum fraction possessed by the thrashed quark has been represented as $x=p^+/P^+$. The Fock-state expansion in the case of two particle for $J^z =\pm1/2$ for the scalar diquark  can be expressed as 
\be
|u~ S\rangle^\pm & =& \int \frac{dx~ d^2\bfp}{2(2\pi)^3\sqrt{x(1-x)}} \Bigg[ \psi^{\pm(\nu)}_{+}(x,\bfp)\bigg|+\frac{1}{2}~s; xP^+,\bfp\bigg\rangle \nonumber \\
 &+& \psi^{\pm(\nu)}_{-}(x,\bfp) \bigg|-\frac{1}{2}~s; xP^+,\bfp\bigg\rangle\Bigg],\label{fockSD}
\ee
where
the flavor index is $\nu=u,d$ and $|\lambda_q~\lambda_S; xP^+,\bfp\rangle$ represents the state of two particle having helicity of thrashed quark as $\lambda_q$ and  helicity of a scalar diquark as $\lambda_S$. In order to differentiate the scalar diquark from the triplet diquark $\lambda_D$, helicity of spin-0 singlet diquark is represented by $\lambda_S=s$.
The LFWFs for the scalar diquark are expressed as \cite{Maji:2017bcz,PRD89,PRD91,majiref24}
\be
\psi^{+(\nu)}_+(x,\bfp)&=& N_S~ \varphi^{(\nu)}_{1}(x,\bfp),\nonumber \\
\psi^{+(\nu)}_-(x,\bfp)&=& N_S\bigg(- \frac{p^1+ip^2}{xM} \bigg)\varphi^{(\nu)}_{2}(x,\bfp),\nonumber \\
\psi^{-(\nu)}_+(x,\bfp)&=& N_S \bigg(\frac{p^1-ip^2}{xM}\bigg) \varphi^{(\nu)}_{2}(x,\bfp),\nonumber \\
\psi^{-(\nu)}_-(x,\bfp)&=&  N_S~ \varphi^{(\nu)}_{1}(x,\bfp).
\label{LFWF_S}
\ee
Here $\varphi^{(\nu)}_i(x,\bfp)$ are LFWFs and $N_S$ is the normalization constant. Similarly, Fock-state expansion in the case of two particle for the vector diquark is given as \cite{majiref25}
\be
|\nu~ A \rangle^\pm & =& \int \frac{dx~ d^2\bfp}{2(2\pi)^3\sqrt{x(1-x)}} \Bigg[ \psi^{\pm(\nu)}_{++}(x,\bfp)\bigg|+\frac{1}{2}~+1; xP^+,\bfp\bigg\rangle \nonumber\\
 &+& \psi^{\pm(\nu)}_{-+}(x,\bfp)\bigg|-\frac{1}{2}~+1; xP^+,\bfp\bigg\rangle +\psi^{\pm(\nu)}_{+0}(x,\bfp)\bigg|+\frac{1}{2}~0; xP^+,\bfp\bigg\rangle \nonumber \\
 &+& \psi^{\pm(\nu)}_{-0}(x,\bfp)\bigg|-\frac{1}{2}~0; xP^+,\bfp\bigg\rangle + \psi^{\pm(\nu)}_{+-}(x,\bfp)\bigg|+\frac{1}{2}~-1; xP^+,\bfp\bigg\rangle \nonumber\\
 &+& \psi^{\pm(\nu)}_{--}(x,\bfp)\bigg|-\frac{1}{2}~-1; xP^+,\bfp\bigg\rangle  \Bigg].\label{fockVD}
\ee
Here $|\lambda_q~\lambda_D; xP^+,\bfp\rangle$ is the state of  two-particle with helicity of quark being $\lambda_q=\pm\frac{1}{2}$ and helicity of vector diquark being $\lambda_D=\pm 1,0$ (triplet). The LFWFs for the vector diquark for the case when $J^z=+1/2$ are given as
\be
\psi^{+(\nu)}_{+~+}(x,\bfp)&=& N^{(\nu)}_1 \sqrt{\frac{2}{3}} \bigg(\frac{p^1-ip^2}{xM}\bigg) \varphi^{(\nu)}_{2}(x,\bfp),\nonumber \\
\psi^{+(\nu)}_{-~+}(x,\bfp)&=& N^{(\nu)}_1 \sqrt{\frac{2}{3}} \varphi^{(\nu)}_{1}(x,\bfp),\nonumber \\
\psi^{+(\nu)}_{+~0}(x,\bfp)&=& - N^{(\nu)}_0 \sqrt{\frac{1}{3}} \varphi^{(\nu)}_{1}(x,\bfp), \nonumber \\
\psi^{+(\nu)}_{-~0}(x,\bfp)&=& N^{(\nu)}_0 \sqrt{\frac{1}{3}} \bigg(\frac{p^1+ip^2}{xM} \bigg)\varphi^{(\nu)}_{2}(x,\bfp),\nonumber \\
\psi^{+(\nu)}_{+~-}(x,\bfp)&=& 0,\nonumber \\
\psi^{+(\nu)}_{-~-}(x,\bfp)&=&  0,
\label{LFWF_Vp}
\ee
and for the case when $J^z=-1/2$, they are given as 
\be
\psi^{-(\nu)}_{+~+}(x,\bfp)&=& 0,\nonumber \\
\psi^{-(\nu)}_{-~+}(x,\bfp)&=& 0,\nonumber \\
\psi^{-(\nu)}_{+~0}(x,\bfp)&=& N^{(\nu)}_0 \sqrt{\frac{1}{3}} \bigg( \frac{p^1-ip^2}{xM} \bigg) \varphi^{(\nu)}_{2}(x,\bfp),\nonumber\\
\psi^{-(\nu)}_{-~0}(x,\bfp)&=& N^{(\nu)}_0\sqrt{\frac{1}{3}} \varphi^{(\nu)}_{1}(x,\bfp),\nonumber \\
\psi^{-(\nu)}_{+~-}(x,\bfp)&=& - N^{(\nu)}_1 \sqrt{\frac{2}{3}} \varphi^{(\nu)}_{1}(x,\bfp),\nonumber \\
\psi^{-(\nu)}_{-~-}(x,\bfp)&=& N^{(\nu)}_1 \sqrt{\frac{2}{3}} \bigg(\frac{p^1+ip^2}{xM}\bigg) \varphi^{(\nu)}_{2}(x,\bfp),
\label{LFWF_Vm}
\ee
where $N_0$, $N_1$ are the normalization constants. Generic ansatz of LFWFs $\varphi^{(\nu)}_i(x,\bfp)$ is being adopted from the soft-wall AdS/QCD prediction \cite{BT,majiref27,PRD89,PRD91} and the parameters $a^\nu_i,~b^\nu_i$ and $\delta^\nu$ are established as \cite{Maji:2017bcz}
\be
\varphi_i^{(\nu)}(x,\bfp)=\frac{4\pi}{\kappa}\sqrt{\frac{\log(1/x)}{1-x}}x^{a_i^\nu}(1-x)^{b_i^\nu}\exp\Bigg[-\delta^\nu\frac{\bfp^2}{2\kappa^2}\frac{\log(1/x)}{(1-x)^2}\bigg].
\label{LFWF_phi}
\ee
The wave functions $\varphi_i^\nu ~(i=1,2)$ reduce to the AdS/QCD prediction for the parameters $a_i^\nu=b_i^\nu=0$  and $\delta^\nu=1.0$. The input parameters of this model which are being used in our calculations, have been discussed in the following section.
%
\section{Input Parameters}\label{secinputp}
The parameters used in our present calculations include $a_i^{\nu}$ and $b_i^{\nu}$, appearing in Eq. \eqref{LFWF_phi}, have been fitted  at model scale $\mu_0$ using the Dirac and Pauli data of form factors \cite{Maji:2016yqo,majiref16,majiref17}. For both the $u$ and $d$ quarks, the parameter $\delta^{\nu}$ is taken as unity at the model scale $\mu_0=0.313~\mathrm{GeV}$. By considering the normalizations in Ref. \cite{Bacchetta:2008af}, the normalized constants $N_{i}^{2}$ in Eqs. (\ref{LFWF_Vp}) and (\ref{LFWF_Vm}) are derived in Ref. \cite{Maji:2016yqo}.
These parameters have been tabulated in Table \ref{tab_par}.
\begin{table}[h]
\centering 
\begin{tabular}{ ||p{1.4cm}||p{1.4cm}|p{1.4cm}|p{1.4cm}|p{1.4cm}|p{1.4cm}||p{1.4cm}|p{1.4cm}||  }
 \hline
 \hline
 ~~$\nu$~~&~~$a_1^{\nu}$~~&~~$b_1^{\nu}$~~&~~$a_2^{\nu}$~~&~~$b_2^{\nu}$~~&~~$\delta^{\nu}$~~&~~$N_0^{\nu}$~~&~~$N_1^{\nu}$~~  \\
 \hline
 \hline
~~$u$~~&~~$0.280$~~&~~$0.1716$~~&~~$0.84$~~&~~$0.2284$~~&~~$1.0$~~&~~$3.2050$~~&~~$0.9895$~~  \\
~~$d$~~&~~$0.5850$~~&~~$0.7000$~~&~~$0.9434$~~&~~$0.64$~~&~~$1.0$~~&~~$5.9423$~~&~~$1.1616$~~    \\
 \hline
 \hline
\end{tabular}
\caption{Values of model parameters corresponding to up and down quarks which appears in Eqs. {\eqref{LFWF_S}}, {\eqref{LFWF_Vp}}, {\eqref{LFWF_Vm}} and {\eqref{LFWF_phi}}.}
\label{tab_par} 
\end{table}

Apart from these, the normalization constant $N_{S}$ in Eq. (\ref{LFWF_S}), corresponding to the isoscalar-scalar diquark, has been determined in Ref. \cite{Maji:2016yqo} and has the value
\begin{equation}
N_{S}=2.0191.
\end{equation}
The coefficients $C_{i}$ of scalar and vector diquarks in  Eq. (\ref{PS_state}) of the proton state has been determined in Ref. \cite{Maji:2016yqo} and given as
\begin{equation}
\begin{aligned}
C_{S}^{2} &=1.3872, \\
C_{V}^{2} &=0.6128, \\
C_{V V}^{2} &=1.
\end{aligned}
\label{Eq3d1}
\end{equation}\
In Eq. (\ref{LFWF_phi}), the AdS/QCD scale parameter $\kappa$ appears. We have used the AdS/QCD scale parameter $\kappa =0.4~\mathrm{GeV}$ as determined in Ref. \cite{majiref28}. The proton mass ($M$) and the constituent quark mass ($m$) are taken to be $0.938~\mathrm{GeV}$ and $0.055~\mathrm{GeV}$ respectively \cite{Chakrabarti:2019wjx}.
\section{Transverse Momentum Dependent Distributions}\label{sectmdsp}
To obtain the TMDs in general, we have to solve the quark-quark correlator.
The unintegrated quark-quark correlator in the light-front formalism for SIDIS is defined as \cite{Maji:2017bcz}
\be
\Phi^{\nu [\Gamma]}(x,\textbf{p}_{\perp};S)&=&\frac{1}{2}\int \frac{dz^- d^2z_T}{2(2\pi)^3} e^{ip.z} \langle P; S_f|\overline{\psi}^\nu (0)\Gamma \mathcal{W}_{[0,z]} \psi^\nu (z) |P;S_i\rangle\Bigg|_{z^+=0}, \label{TMDcor}
\ee
at equal light-front time $z^+=0$. Thrashed quark's longitudinal momentum fraction is $x=p^+/P^+$ and its helicity is $\lambda$. Proton's momentum is denoted by $P$ and its heicity is $\lambda_N$. Light-cone gauge $A^+=0$ is selected and a frame is chosen where the momentum of the proton is $ P\equiv (P^+,\frac{M^2}{P^+},\textbf{0} ),$ the momentum of virtual photon is $q\equiv (x_B P^+, \frac{Q^2}{x_BP^+},\textbf{0})$, where $x_B= \frac{Q^2}{2P.q}$ is the Bjorken variable and $Q^2 = -q^2$. If the helicity of proton is $\lambda_N$ then its spin components are written by $S^+ = \lambda_N \frac{P^+}{M},~ S^- = \lambda_N\frac{P^-}{M},$ and $ S_T $. In our study, we have taken the value of Wilson line to be $1$. 
In twist-4 case, there are total of 8 TMDs out of which $6$ are T-even and  $2$ are T-odd. For different values of $(\Gamma)$, the twist-4 quark TMDs are projected in the form  Eq. (\ref{TMDcor}) following Ref. \cite{liu21} and are expressed as
\begin{eqnarray}
\Phi^{\nu[\gamma^-]}&=&\frac{M^2}{(P^+)^2}[{\color{blue}f_3}-\frac{\epsilon_T^{ij}\bm{p}_{Ti}\bm{S}_{Tj}}{M}{\color{red}f_{3T}^{\perp}}],\label{eqtmdlist1}\\
\Phi^{\nu[\gamma^-\gamma_5]}&=&\frac{M^2}{(P^+)^2}[\lambda {\color{blue}g_{3L}}+\frac{\bm{p}_T\cdot\bm{S}_T}{M}{\color{blue}g_{3T}}],   \label{eqtmdlist2}\\
\Phi^{\nu[\bm{i}\sigma^{i-}\gamma_5]}&=&\frac{M^2}{(P^+)^2}[\bm{S}_{T}^i{\color{blue}h_{3T}}+\frac{\bm{p}_{T}^i}{M}(\lambda {\color{blue}h_{3L}^{\perp}}+\frac{\bm{p}_T\cdot\bm{S}_T}{M}{\color{blue}h_{3T}^{\perp})}-\frac{\epsilon_T^{ij}\bm{p}_{Tj}}{M}{\color{red}h_3^{\perp}}]\label{eqtmdlist3}\\
                                     &=&\frac{M^2}{(P^+)^2}[\bm{S}_{T}^i{\color{blue}h_3}+\lambda\frac{\bm{p}_{T}^i}{M}{\color{blue}h_{3L}^{\perp}}+\frac{(\bm{p}_T^i\bm{p}_{T}^j-\frac{1}{2}\bm{p}_T^2g_T^{ij})\bm{S}_{Tj}}{M^2}{\color{blue}h_{3T}^{\perp}}-\frac{\epsilon_T^{ij}\bm{p}_{Tj}}{M}{\color{red}h_3^{\perp}}],   \label{eqtmdlist4}                        
\end{eqnarray}
where, $h_3=h_{3T}+\frac{\bm{p}_T^2}{2M^2}h_{3T}^{\perp}$. T-even and T-odd TMDs are represented in blue and red color sequentially (color online). In this work, our focus is on twist-4 T-even TMDs. From Eqs. {\eqref{eqtmdlist1}-\eqref{eqtmdlist4}} it is understood that $f_3 =f_3^{\nu}(x,\textbf{p}_{\perp}^2) $ and similarly for the other TMDs. We have used the  standard notation for  $\sigma^{k l}=i\left[\gamma^{k}, \gamma^{l}\right] / 2$, $\kappa^{jk}=(p_T^jp_T^k-\frac12\delta^{jk}p_T^2)$ where $p_T^2=|\vec{p}_T|^2$. We have used the definition $\varepsilon_{T}^{i j}=\varepsilon^{-+i j}$, where $\varepsilon^{23}=-\varepsilon^{32}=1$ and it is zero when  $i,\,j$ are same. Transverse directions are being designated by the indices $i$ and $j$.
\section{Results}\label{secresults}
\subsection{Overlap Form}
It is convenient to express TMDs in the overlap form of wave functions detailing the initial and final state spin of the quarks and proton. To get this form for the scalar diquark, we have to substitute Eq. \eqref{fockSD} with suitable polarization in Eq. \eqref{TMDcor} via Eq. \eqref{PS_state}. After taking a particular correlation (for $\Gamma= \gamma^-,  \gamma^-\gamma_5$ and $\sigma^{i-}\gamma_5$) from Eqs. {\eqref{eqtmdlist1}-\eqref{eqtmdlist4}}, we can compute a particular TMD by taking the appropriate combination of the polarization of proton. For example, to calculate the
unpolarized TMD $f_{3}^{\nu}(x, {\bf p_\perp^2})$ appearing in Eq. \eqref{eqtmdlist1}, we have to use $(\Gamma=\gamma^-)$ and spin of the nucleon has to be taken as  $+1/2$ in both the initial and final states. In this way, TMDs in terms of LFWFs for the scalar diquark can be expressed  as
\be
x^2 f_{3}^{\nu(S)}(x, {\bf p_\perp^2}) &=&  \frac{1}{16 \pi^3} \bigg(\frac{p_\perp^2+m^2}{M^2}\bigg)\Bigg[|\psi ^{+\nu}_+(x,\textbf{p}_{\perp})|^2+|\psi ^{ + \nu}_-(x,\textbf{p}_{\perp})|^2\Bigg], \label{oef3s}
%
\ee
\be
x^2 g_{3L}^{\nu(S)}(x, {\bf p_\perp^2}) &=&  \frac{1}{32 \pi^3 M^2} \Bigg[\big({p_\perp^2-m^2}\big)\bigg[|\psi ^{+\nu}_+(x,\textbf{p}_{\perp})|^2-|\psi ^{ + \nu}_-(x,\textbf{p}_{\perp})|^2\bigg] \nonumber\\
&& + 2m ({\textbf{p}_{x}}-\iota {\textbf{p}_{y}})\bigg[\psi ^{+\nu \dagger}_+(x,\textbf{p}_{\perp})\psi^{+ \nu}_-(x,\textbf{p}_{\perp})\bigg] \nonumber \\
&&+ 2m ({\textbf{p}_{x}}+\iota {\textbf{p}_{y}}) \bigg[\psi ^{+\nu \dagger}_-(x,\textbf{p}_{\perp})\psi^{+ \nu}_+(x,\textbf{p}_{\perp})\bigg]\Bigg], \label{oeg3ls}
\\
%
%
x^2 {\textbf{p}_{x}}h_{3L}^{\perp\nu(S)}(x, {\bf p_\perp^2}) &=&  \frac{1}{32 \pi^3 M} \Bigg[2m{\textbf{p}_{x}} \bigg[|\psi ^{+\nu}_+(x,\textbf{p}_{\perp})|^2-|\psi ^{ + \nu}_-(x,\textbf{p}_{\perp})|^2\bigg] \nonumber\\
&& + \bigg( m^2 -({\textbf{p}_{x}}-\iota {\textbf{p}_{y}})^2 \bigg)
\bigg[\psi ^{+\nu \dagger}_+(x,\textbf{p}_{\perp})\psi^{+ \nu}_-(x,\textbf{p}_{\perp})\bigg] \nonumber \\
&&+ \bigg( m^2 -({\textbf{p}_{x}}+\iota {\textbf{p}_{y}})^2 \bigg) \bigg[\psi ^{+\nu \dagger}_-(x,\textbf{p}_{\perp})\psi^{+ \nu}_+(x,\textbf{p}_{\perp})\bigg]\Bigg]. \label{oeh3lps}\\
%
%
x^2 {\textbf{p}_{x}} g_{3T}^{\nu(S)}(x, {\bf p_\perp^2}) &=&  \frac{1}{32 \pi^3 M} \Bigg[\big({p_\perp^2-m^2}\big)\bigg[
\psi ^{+\nu \dagger}_+(x,\textbf{p}_{\perp})\psi^{- \nu}_+(x,\textbf{p}_{\perp}) - \psi ^{+\nu \dagger}_-(x,\textbf{p}_{\perp}) \psi ^{-\nu}_-(x,\textbf{p}_{\perp}) \nonumber\\
 &&+ \psi ^{-\nu \dagger}_+(x,\textbf{p}_{\perp}) \psi ^{+\nu}_+(x,\textbf{p}_{\perp})-\psi^{- \dagger \nu}_-(x,\textbf{p}_{\perp}) \psi ^{+\nu}_-(x,\textbf{p}_{\perp})\bigg] \nonumber\\
&& + 2m ({\textbf{p}_{x}}-\iota {\textbf{p}_{y}})\bigg[\psi ^{+\nu \dagger}_+(x,\textbf{p}_{\perp})\psi^{- \nu}_-(x,\textbf{p}_{\perp}) + \psi ^{-\nu \dagger}_+(x,\textbf{p}_{\perp})\psi^{+ \nu}_-(x,\textbf{p}_{\perp})\bigg] \nonumber \\
&&+ 2m ({\textbf{p}_{x}}+\iota {\textbf{p}_{y}}) \bigg[\psi ^{+\nu \dagger}_-(x,\textbf{p}_{\perp})\psi^{- \nu}_+(x,\textbf{p}_{\perp}) + \psi ^{-\nu \dagger}_-(x,\textbf{p}_{\perp})\psi^{+ \nu}_+(x,\textbf{p}_{\perp})\bigg]\Bigg], \label{oeg3ts} \\
%
%
x^2 \bigg[h_{3T}^{\nu(S)}(x, {\bf p_\perp^2})+\frac{{\textbf{p}_{x}}^2}{M^2}h_{3T}^{\perp\nu(S)}(x, {\bf p_\perp^2})\bigg]
&=&  \frac{1}{32 \pi^3 M^2} \Bigg[2m{\textbf{p}_{x}} \bigg[\psi ^{+\nu \dagger}_+(x,\textbf{p}_{\perp})\psi^{- \nu}_+(x,\textbf{p}_{\perp}) - \psi ^{+\nu \dagger}_-(x,\textbf{p}_{\perp}) \psi ^{-\nu}_-(x,\textbf{p}_{\perp}) \nonumber\\
 &&+ \psi ^{-\nu \dagger}_+(x,\textbf{p}_{\perp}) \psi ^{+\nu}_+(x,\textbf{p}_{\perp})-\psi^{- \dagger \nu}_-(x,\textbf{p}_{\perp}) \psi ^{+\nu}_-(x,\textbf{p}_{\perp})\bigg] \nonumber\\
&& + \bigg( m^2 -({\textbf{p}_{x}}-\iota {\textbf{p}_{y}})^2 \bigg)
\bigg[\psi ^{+\nu \dagger}_+(x,\textbf{p}_{\perp})\psi^{- \nu}_-(x,\textbf{p}_{\perp}) + \psi ^{-\nu \dagger}_+(x,\textbf{p}_{\perp})\psi^{+ \nu}_-(x,\textbf{p}_{\perp})\bigg] \nonumber \\
&&+ \bigg( m^2 -({\textbf{p}_{x}}+\iota {\textbf{p}_{y}})^2 \bigg) \bigg[\psi ^{+\nu \dagger}_-(x,\textbf{p}_{\perp})\psi^{- \nu}_+(x,\textbf{p}_{\perp}) + \psi ^{-\nu \dagger}_-(x,\textbf{p}_{\perp})\psi^{+ \nu}_+(x,\textbf{p}_{\perp})\bigg]\Bigg], \nonumber\\
\label{oeh3ts}\\
%
%
x^2 {\textbf{p}_{x}}{\textbf{p}_{y}}h_{3T}^{\perp\nu(S)}(x, {\bf p_\perp^2}) &=&  \frac{1}{32 \pi^3} \Bigg[2m{\textbf{p}_{y}} \bigg[\psi ^{+\nu \dagger}_+(x,\textbf{p}_{\perp})\psi^{- \nu}_+(x,\textbf{p}_{\perp}) - \psi ^{+\nu \dagger}_-(x,\textbf{p}_{\perp}) \psi ^{-\nu}_-(x,\textbf{p}_{\perp}) \nonumber\\
 &&+ \psi ^{-\nu \dagger}_+(x,\textbf{p}_{\perp}) \psi ^{+\nu}_+(x,\textbf{p}_{\perp}-\psi^{- \dagger \nu}_-(x,\textbf{p}_{\perp}) \psi ^{+\nu}_-(x,\textbf{p}_{\perp})\bigg] \nonumber\\
&& -\iota \bigg( m^2 +({\textbf{p}_{x}}-\iota {\textbf{p}_{y}})^2 \bigg)
\bigg[\psi ^{+\nu \dagger}_+(x,\textbf{p}_{\perp})\psi^{- \nu}_-(x,\textbf{p}_{\perp}) + \psi ^{-\nu \dagger}_+(x,\textbf{p}_{\perp})\psi^{+ \nu}_-(x,\textbf{p}_{\perp})\bigg] \nonumber \\
&&+\iota \bigg( m^2 +({\textbf{p}_{x}}+\iota {\textbf{p}_{y}})^2 \bigg) \bigg[\psi ^{+\nu \dagger}_-(x,\textbf{p}_{\perp})\psi^{- \nu}_+(x,\textbf{p}_{\perp}) + \psi ^{-\nu \dagger}_-(x,\textbf{p}_{\perp})\psi^{+ \nu}_+(x,\textbf{p}_{\perp})\bigg]\Bigg].\nonumber\\
\label{oeh3tps}
\ee
Further, to get overlap form of TMDs for the vector diquark, we have to substitute Eq. \eqref{fockVD} with suitable polarization in Eq. \eqref{TMDcor} via Eq. \eqref{PS_state}. TMDs in terms of LFWFs for the vector diquark can be written  as
\be
%
%
x^2 f_{3}^{\nu(A)}(x, {\bf p_\perp^2}) &=&\sum_{\lambda_A}  \frac{1}{16 \pi^3} \bigg(\frac{p_\perp^2+m^2}{M^2}\bigg)\Bigg[|\psi ^{+\nu}_{+\lambda_A}(x,\textbf{p}_{\perp})|^2+|\psi ^{ + \nu}_{-\lambda_A}(x,\textbf{p}_{\perp})|^2\Bigg], \label{oef3v}\\
%
%
x^2 g_{3L}^{\nu(A)}(x, {\bf p_\perp^2}) &=&\sum_{\lambda_A}  \frac{1}{32 \pi^3 M^2} \Bigg[\big({p_\perp^2-m^2}\big)\bigg[|\psi ^{+\nu}_{+\lambda_A}(x,\textbf{p}_{\perp})|^2-|\psi ^{ + \nu}_{-\lambda_A}(x,\textbf{p}_{\perp})|^2\bigg] \nonumber\\
&& + 2m ({\textbf{p}_{x}}-\iota {\textbf{p}_{y}})\bigg[\psi ^{+\nu \dagger}_{+\lambda_A}(x,\textbf{p}_{\perp})\psi^{+ \nu}_{-\lambda_A}(x,\textbf{p}_{\perp})\bigg] \nonumber \\
&&+ 2m ({\textbf{p}_{x}}+\iota {\textbf{p}_{y}}) \bigg[\psi ^{+\nu \dagger}_{-\lambda_A}(x,\textbf{p}_{\perp})\psi^{+ \nu}_{+\lambda_A}(x,\textbf{p}_{\perp})\bigg]\Bigg], \label{oeg3lpv} \\
%
%
x^2 {\textbf{p}_{x}}h_{3L}^{\perp\nu(A)}(x, {\bf p_\perp^2}) &=&\sum_{\lambda_A}  \frac{1}{32 \pi^3 M} \Bigg[2m{\textbf{p}_{x}} \bigg[|\psi ^{+\nu}_{+\lambda_A}(x,\textbf{p}_{\perp})|^2-|\psi ^{ + \nu}_{-\lambda_A}(x,\textbf{p}_{\perp})|^2\bigg] \nonumber\\
&& + \bigg( m^2 -({\textbf{p}_{x}}-\iota {\textbf{p}_{y}})^2 \bigg)
\bigg[\psi ^{+\nu \dagger}_{+\lambda_A}(x,\textbf{p}_{\perp})\psi^{+ \nu}_{-\lambda_A}(x,\textbf{p}_{\perp})\bigg] \nonumber \\
&&+ \bigg( m^2 -({\textbf{p}_{x}}+\iota {\textbf{p}_{y}})^2 \bigg) \bigg[\psi ^{+\nu \dagger}_{-\lambda_A}(x,\textbf{p}_{\perp})\psi^{+ \nu}_{+\lambda_A}(x,\textbf{p}_{\perp})\bigg]\Bigg], \label{oeh3lpv}\\
%
%
x^2 {\textbf{p}_{x}} g_{3T}^{\nu(A)}(x, {\bf p_\perp^2}) &=&  \frac{1}{32 \pi^3 M} \Bigg[\big({p_\perp^2-m^2}\big)\bigg[
\psi ^{+\nu \dagger}_{+0}(x,\textbf{p}_{\perp})\psi^{- \nu}_{+0}(x,\textbf{p}_{\perp}) - \psi ^{+\nu \dagger}_{-0}(x,\textbf{p}_{\perp}) \psi ^{-\nu}_{-0}(x,\textbf{p}_{\perp}) \nonumber\\
 &&+ \psi ^{-\nu \dagger}_{+0}(x,\textbf{p}_{\perp}) \psi ^{+\nu}_{+0}(x,\textbf{p}_{\perp})-\psi^{- \dagger \nu}_{-0}(x,\textbf{p}_{\perp}) \psi ^{+\nu}_{-0}(x,\textbf{p}_{\perp})\bigg] \nonumber\\
&& + 2m ({\textbf{p}_{x}}-\iota {\textbf{p}_{y}})\bigg[\psi ^{+\nu \dagger}_{+0}(x,\textbf{p}_{\perp})\psi^{- \nu}_{-0}(x,\textbf{p}_{\perp}) + \psi ^{-\nu \dagger}_{+0}(x,\textbf{p}_{\perp})\psi^{+ \nu}_{-0}(x,\textbf{p}_{\perp})\bigg] \nonumber \\
&&+ 2m ({\textbf{p}_{x}}+\iota {\textbf{p}_{y}}) \bigg[\psi ^{+\nu \dagger}_{-0}(x,\textbf{p}_{\perp})\psi^{- \nu}_{+0}(x,\textbf{p}_{\perp}) + \psi ^{-\nu \dagger}_{-0}(x,\textbf{p}_{\perp})\psi^{+ \nu}_{+0}(x,\textbf{p}_{\perp})\bigg]\Bigg], \label{oeg3tv} \\
%
%
x^2 \bigg[h_{3T}^{\nu(A)}(x, {\bf p_\perp^2})+\frac{{\textbf{p}_{x}}^2}{M^2}h_{3T}^{\perp\nu(A)}(x, {\bf p_\perp^2})\bigg]
&=&  \frac{1}{32 \pi^3 M^2} \Bigg[2m{\textbf{p}_{x}} \bigg[\psi ^{+\nu \dagger}_{+0}(x,\textbf{p}_{\perp})\psi^{- \nu}_{+0}(x,\textbf{p}_{\perp}) - \psi ^{+\nu \dagger}_{-0}(x,\textbf{p}_{\perp}) \psi ^{-\nu}_{-0}(x,\textbf{p}_{\perp}) \nonumber\\
 &&+ \psi ^{-\nu \dagger}_{+0}(x,\textbf{p}_{\perp}) \psi ^{+\nu}_{+0}(x,\textbf{p}_{\perp})-\psi^{- \dagger \nu}_{-0}(x,\textbf{p}_{\perp}) \psi ^{+\nu}_{-0}(x,\textbf{p}_{\perp})\bigg] \nonumber\\
&& + \bigg( m^2 -({\textbf{p}_{x}}-\iota {\textbf{p}_{y}})^2 \bigg)
\bigg[\psi ^{+\nu \dagger}_{+0}(x,\textbf{p}_{\perp})\psi^{- \nu}_{-0}(x,\textbf{p}_{\perp}) + \psi ^{-\nu \dagger}_{+0}(x,\textbf{p}_{\perp})\psi^{+ \nu}_{-0}(x,\textbf{p}_{\perp})\bigg] \nonumber \\
&&+ \bigg( m^2 -({\textbf{p}_{x}}+\iota {\textbf{p}_{y}})^2 \bigg) \bigg[\psi ^{+\nu \dagger}_{-0}(x,\textbf{p}_{\perp})\psi^{- \nu}_{+0}(x,\textbf{p}_{\perp}) + \psi ^{-\nu \dagger}_{-0}(x,\textbf{p}_{\perp})\psi^{+ \nu}_{+0}(x,\textbf{p}_{\perp})\bigg]\Bigg], \nonumber\\
\label{oeh3tv}\\
%
%
x^2 {\textbf{p}_{x}}{\textbf{p}_{y}}h_{3T}^{\perp\nu(A)}(x, {\bf p_\perp^2}) &=&  \frac{1}{32 \pi^3} \Bigg[2m{\textbf{p}_{y}} \bigg[\psi ^{+\nu \dagger}_{+0}(x,\textbf{p}_{\perp})\psi^{- \nu}_{+0}(x,\textbf{p}_{\perp}) - \psi ^{+\nu \dagger}_{-0}(x,\textbf{p}_{\perp}) \psi ^{-\nu}_{-0}(x,\textbf{p}_{\perp}) \nonumber\\
 &&+ \psi ^{-\nu \dagger}_{+0}(x,\textbf{p}_{\perp}) \psi ^{+\nu}_{+0}(x,\textbf{p}_{\perp}-\psi^{- \dagger \nu}_{-0}(x,\textbf{p}_{\perp}) \psi ^{+\nu}_{-0}(x,\textbf{p}_{\perp})\bigg] \nonumber\\
&& -\iota \bigg( m^2 +({\textbf{p}_{x}}-\iota {\textbf{p}_{y}})^2 \bigg)
\bigg[\psi ^{+\nu \dagger}_{+0}(x,\textbf{p}_{\perp})\psi^{- \nu}_{-0}(x,\textbf{p}_{\perp}) + \psi ^{-\nu \dagger}_{+0}(x,\textbf{p}_{\perp})\psi^{+ \nu}_{-0}(x,\textbf{p}_{\perp})\bigg] \nonumber \\
&&+\iota \bigg( m^2 +({\textbf{p}_{x}}+\iota {\textbf{p}_{y}})^2 \bigg) \bigg[\psi ^{+\nu \dagger}_{-0}(x,\textbf{p}_{\perp})\psi^{- \nu}_{+0}(x,\textbf{p}_{\perp}) + \psi ^{-\nu \dagger}_{-0}(x,\textbf{p}_{\perp})\psi^{+ \nu}_{+0}(x,\textbf{p}_{\perp})\bigg]\Bigg],\nonumber\\
\label{oeh3tpv}
\ee
where $\lambda_A$ runs over $0,\pm$, which is nothing but the summation over vector diquark's helicity.

\subsection{Explicit Expressions of TMDs}
By substituting the expressions of LFWFs for the scalar diquark from Eq. (\ref{LFWF_S}) into Eqs. {\eqref{oef3s}-\eqref{oeh3tps}}, we have obtained the explicit expressions of twist-4 T-even TMDs $f_{3}^{\nu}(x, {\bf p_\perp^2}),~ g_{3L}^{\nu}(x, {\bf p_\perp^2}),$ $h_{3L}^{\perp\nu}(x, {\bf p_\perp^2}),$ $g_{3T}^{\nu}(x, {\bf p_\perp^2}),~h_{3T}^{\nu}(x, {\bf p_\perp^2})$ and $h_{3T}^{\perp \nu}(x, {\bf p_\perp^2})$ for the scalar diquark and are expressed as
\be
%
x^2 f_{3}^{\nu(S)}(x, {\bf p_\perp^2}) &=&  \frac{C_{S}^{2} N_s^2}{16 \pi^3} \bigg(\frac{p_\perp^2+m^2}{M^2}\bigg) \Bigg[|\varphi_1^\nu|^2 + \frac{p_\perp^2}{x^2 M^2}|\varphi_2^\nu|^2\Bigg], \label{eef3s} \\
%
%
x^2 g_{3L}^{\nu(S)}(x, {\bf p_\perp^2}) &=&  \frac{C_{S}^{2} N_s^2}{32 \pi^3 M^2} \Bigg[\big({p_\perp^2-m^2}\big)\bigg[|\varphi_1^\nu|^2 - \frac{p_\perp^2}{x^2 M^2}|\varphi_2^\nu|^2\bigg]-\frac{4m {p_\perp^2}}{x M}|\varphi_1^\nu||\varphi_2^\nu| \Bigg],  \label{eeg3ls} \\
%
%
x^2 h_{3L}^{\perp\nu(S)}(x, {\bf p_\perp^2}) &=&  \frac{C_{S}^{2} N_s^2}{16 \pi^3 M} \Bigg[m\bigg[|\varphi_1^\nu|^2 - \frac{p_\perp^2}{x^2 M^2}|\varphi_2^\nu|^2\bigg]+\frac{\big({p_\perp^2-m^2}\big)}{x M}|\varphi_1^\nu||\varphi_2^\nu| \Bigg], \label{eeh3lps}\\ 
%
%
%
x^2 g_{3T}^{\nu(S)}(x, {\bf p_\perp^2}) &=&  \frac{C_{S}^{2} N_s^2}{8 \pi^3 M} \Bigg[m\bigg[|\varphi_1^\nu|^2 - \frac{p_\perp^2}{x^2 M^2}|\varphi_2^\nu|^2\bigg]+\frac{\big({p_\perp^2-m^2}\big)}{x M}|\varphi_1^\nu||\varphi_2^\nu| \Bigg], \label{eeg3ts} \\
%
%
x^2 h_{3T}^{\nu(S)}(x, {\bf p_\perp^2})&=&  \frac{C_{S}^{2} N_s^2}{16 \pi^3} \bigg(\frac{p_\perp^2+m^2}{M^2}\bigg) \Bigg[|\varphi_1^\nu|^2 + \frac{p_\perp^2}{x^2 M^2}|\varphi_2^\nu|^2\Bigg], \label{eeh3ts} \\
%
x^2 h_{3T}^{\perp \nu(S)}(x, {\bf p_\perp^2}) &=&  -\frac{C_{S}^{2} N_s^2}{8 \pi^3} \Bigg[|\varphi_1^\nu|^2 + \frac{m^2}{x^2 M^2}|\varphi_2^\nu|^2-\frac{2m}{x M}|\varphi_1^\nu||\varphi_2^\nu| \Bigg]. \label{eeh3tps}
\ee
Similarly, by substituting the expressions of LFWFs for the vector diquark from Eqs. (\ref{LFWF_Vp}) and (\ref{LFWF_Vm}) into Eqs. {\eqref{oef3v}-\eqref{oeh3tpv}}, we have obtained the explicit expressions of twist-4 T-even TMDs $f_{3}^{\nu}(x, {\bf p_\perp^2}),~ g_{3L}^{\nu}(x, {\bf p_\perp^2}),~h_{3L}^{\perp\nu}(x, {\bf p_\perp^2}),~g_{3T}^{\nu}(x, {\bf p_\perp^2}),~h_{3T}^{\nu}(x, {\bf p_\perp^2})$ and $h_{3T}^{\perp \nu}(x, {\bf p_\perp^2})$ for the vector diquark and are expressed as
\be
%
%
   x^2 f_{3}^{\nu(A)}(x, {\bf p_\perp^2}) &=& \frac{C_{A}^{2}}{16 \pi^3}  \bigg(\frac{1}{3} |N_0^\nu|^2+\frac{2}{3}|N_1^\nu|^2 \bigg)\bigg(\frac{p_\perp^2+m^2}{M^2}\bigg) \Bigg[|\varphi_1^\nu|^2 + \frac{p_\perp^2}{x^2 M^2}|\varphi_2^\nu|^2\Bigg], \label{eef3v}\\
%
%
x^2 g_{3L}^{\nu(A)}(x, {\bf p_\perp^2}) &=& \frac{C_{A}^{2}}{32 \pi^3 M^2}  \bigg(\frac{1}{3} |N_0^\nu|^2-\frac{2}{3}|N_1^\nu|^2 \bigg) \Bigg[\big({p_\perp^2-m^2}\big)\bigg[|\varphi_1^\nu|^2 - \frac{p_\perp^2}{x^2 M^2}|\varphi_2^\nu|^2\bigg]\nonumber \\
&&-\frac{4m {p_\perp^2}}{x M}|\varphi_1^\nu||\varphi_2^\nu| \Bigg],  \label{eeg3lv}\\
%
%
x^2 h_{3L}^{\perp\nu(A)}(x, {\bf p_\perp^2}) &=& \frac{C_{A}^{2}}{16 \pi^3 M}  \bigg(\frac{1}{3} |N_0^\nu|^2-\frac{2}{3}|N_1^\nu|^2 \bigg) \Bigg[m\bigg[|\varphi_1^\nu|^2 - \frac{p_\perp^2}{x^2 M^2}|\varphi_2^\nu|^2\bigg]+\frac{\big({p_\perp^2-m^2}\big)}{x M}|\varphi_1^\nu||\varphi_2^\nu| \Bigg], \label{eeh3lpv}\\
%
%
%
x^2 g_{3T}^{\nu(A)}(x, {\bf p_\perp^2}) &=&  -\frac{C_{A}^{2}}{8 \pi^3 M}\bigg(\frac{1}{3} |N_0^\nu|^2\bigg) \Bigg[m\bigg[|\varphi_1^\nu|^2 - \frac{p_\perp^2}{x^2 M^2}|\varphi_2^\nu|^2\bigg]+\frac{\big({p_\perp^2-m^2}\big)}{x M}|\varphi_1^\nu||\varphi_2^\nu| \Bigg],  \label{eeg3tv} \\
%
%
x^2 h_{3T}^{\nu(A)}(x, {\bf p_\perp^2})&=&  -\frac{C_{A}^{2}}{16 \pi^3}\bigg(\frac{1}{3} |N_0^\nu|^2\bigg) \bigg(\frac{p_\perp^2+m^2}{M^2}\bigg) \Bigg[|\varphi_1^\nu|^2 + \frac{p_\perp^2}{x^2 M^2}|\varphi_2^\nu|^2\Bigg], \label{eeh3tv} 
\ee
\be
%
x^2 h_{3T}^{\perp \nu(A)}(x, {\bf p_\perp^2}) &=&  \frac{C_{A}^{2}}{8 \pi^3}\bigg(\frac{1}{3} |N_0^\nu|^2\bigg) \Bigg[|\varphi_1^\nu|^2 + \frac{m^2}{x^2 M^2}|\varphi_2^\nu|^2-\frac{2m}{x M}|\varphi_1^\nu||\varphi_2^\nu| \Bigg].  \label{eeh3tpv}
\ee
\section{Discussion}\label{secdiscussion}
\subsection{Relation among TMDs in LFQDM}
As discussed earlier, very less work has been done on twist-4 T-even TMDs whereas on the other hand, leading and subleading twist T-even TMDs have been explored quite well. Also model dependent/independent QCD relations between leading, subleading and twist-4 TMDs have been explored \cite{Lorce:2014hxa,Avakian:2010br,Bastami:2020rxn,majiref16,liu21}.
The relations between the different twist-4 TMDs  in the same model are
\be
x^2~{f}^{\nu(S)}_3(x, {\bf p_\perp^2}) \overset{LFQDM}{=}& x^2~{h}^{\nu(S)}_{3T}(x, {\bf p_\perp^2}),
\label{cef3b}\\
x^2~{g}^{\nu(S)}_{3T}(x, {\bf p_\perp^2})\overset{LFQDM}{=}&2 x^2~{h}^{\perp\nu(S)}_{3L}(x, {\bf p_\perp^2}).
\label{ceg3tb}
\ee
The above expression for ${g}^{\nu(S)}_{3T}(x, {\bf p_\perp^2})$ also holds in the spectator model \cite{liu21}.

For better understanding of our results of twist-4 T-even TMDs, we have expressed them in the form of available leading twist T-even TMDs \cite{Maji:2017bcz,Maji:2016yqo} within the same model. The unpolarized twist-4 T-even TMD $f_{3}^{\nu}(x, {\bf p_\perp^2})$ can be expressed in the form of unpolarized T-even leading twist TMD ${f}^{\nu  }_1(x, {\bf p_\perp^2})$ as
\be
x^2~{f}^{\nu  }_3(x, {\bf p_\perp^2}) \overset{LFQDM}{=}&\bigg(~\frac{p_\perp^2+m^2}{M^2}\bigg)~{f}^{\nu  }_1(x, {\bf p_\perp^2}).
\label{cef3}
\ee
This expression has also been obtained in Ref. \cite{Lorce:2014hxa} with an additional tilde term on right hand side, for the case of an ensemble of free quarks. After solving the quark-quark correlator in Eq. \eqref{TMDcor} for the extraction of ${f}^{\nu  }_1(x, {\bf p_\perp^2})$ \cite{Maji:2017bcz} and ${f}^{\nu  }_3(x, {\bf p_\perp^2})$, we have discovered that tilde term appearing in Ref. \cite{Lorce:2014hxa} disappear, since our model has no quark-model interactions. The expressions of longitudinally polarized twist-4 T-even TMDs $\bigg(g_{3L}^{\nu}(x, {\bf p_\perp^2})$ and $ h_{3L}^{\perp\nu}(x, {\bf p_\perp^2})\bigg)$ can be written in the form of leading twist T-even longitudinally polarized TMDs  $\bigg({h}^{\nu\perp}_{1L}(x, {\bf p_\perp^2})$ and ${g}^{\nu  }_{1L}(x, {\bf p_\perp^2})\bigg)$ as
\be
x^2~{g}^{\nu  }_{3L}(x, {\bf p_\perp^2})\overset{LFQDM}{=}& \bigg(~\frac{p_\perp^2-m^2}{M^2}\bigg)~{g}^{\nu  }_{1L}(x, {\bf p_\perp^2})+\frac{2 m p_\perp^2}{M^3}{h}^{\nu\perp}_{1L}(x, {\bf p_\perp^2}),
\label{ceg3l}\\
x^2~{h}^{\nu\perp}_{3L}(x, {\bf p_\perp^2})\overset{LFQDM}{=}&-\bigg(~\frac{p_\perp^2-m^2}{M^2}\bigg)~{h}^{\nu\perp}_{1L}(x, {\bf p_\perp^2})+\frac{2 m}{M}{g}^{\nu}_{1L}(x, {\bf p_\perp^2}).
\label{ceh3lp}
\ee
The transversely polarized twist-4 T-even TMDs \bigg(${g}^{\nu  }_{3T}(x, {\bf p_\perp^2}),~{h}^{\nu  }_{3T}(x, {\bf p_\perp^2})$ and ${h}^{\nu\perp}_{3T}(x, {\bf p_\perp^2})$\bigg) can be expressed in the form of transversely polarized leading twist T-even TMDs \bigg(${g}^{\nu  }_{1T}(x, {\bf p_\perp^2}),$ ${h}^{\nu}_{1T}(x, {\bf p_\perp^2}),~{h}^{\nu\perp}_{1T}(x, {\bf p_\perp^2}) $ and ${h}^{\nu}_{1}(x, {\bf p_\perp^2})$\bigg) as
\be
x^2~{g}^{\nu  }_{3T}(x, {\bf p_\perp^2})&\overset{LFQDM}{=}&\bigg(~\frac{p_\perp^2-m^2}{M^2}\bigg)~{g}^{\nu  }_{1T}(x, {\bf p_\perp^2})+\frac{m p_\perp^2}{M^3}{h}^{\nu\perp}_{1T}(x, {\bf p_\perp^2})+\frac{2m}{M}~{h}^{\nu}_{1}(x, {\bf p_\perp^2}),
\label{ceg3t}\\
x^2~{h}^{\nu  }_{3T}(x, {\bf p_\perp^2})&\overset{LFQDM}{=}&\bigg(~\frac{p_\perp^2+m^2}{M^2}\bigg)~{h}^{\nu  }_{1T}(x, {\bf p_\perp^2}),
\label{ceh3t}\\
x^2~{h}^{\nu\perp}_{3T}(x, {\bf p_\perp^2})&\overset{LFQDM}{=}&\frac{m^2}{M^2}{h}^{\nu\perp}_{1T}(x, {\bf p_\perp^2})-{2}~{h}^{\nu}_{1}(x, {\bf p_\perp^2})
+\frac{2m}{M}~{g}^{\nu  }_{1T}(x, {\bf p_\perp^2}).
\label{ceh3tp}
\ee
In future work, it would be interesting to find the model relations of twist-4 TMDs with subleading twist.
\begin{figure*}
\centering
\begin{minipage}[c]{0.98\textwidth}
(a)\includegraphics[width=7.5cm]{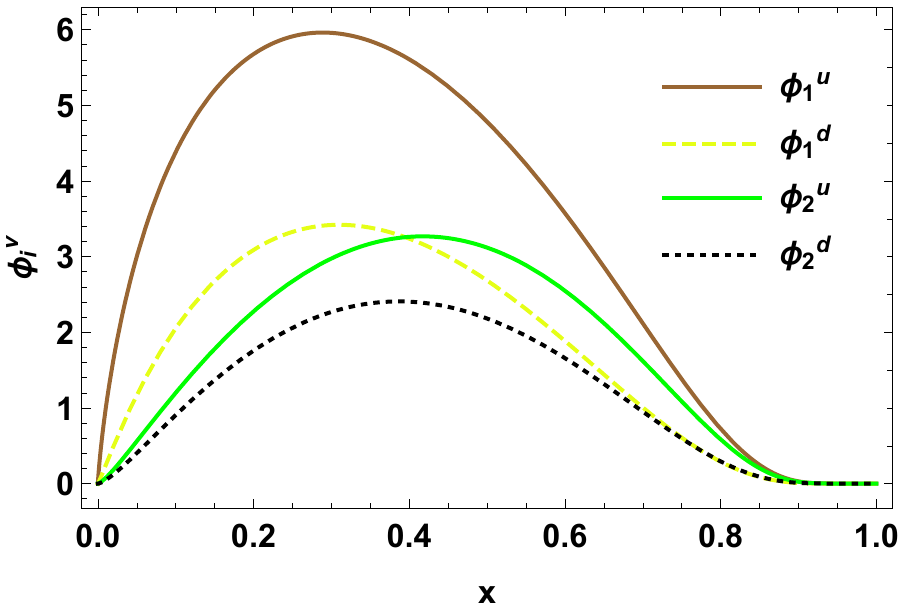}
\hspace{0.05cm}
(b)\includegraphics[width=7.5cm]{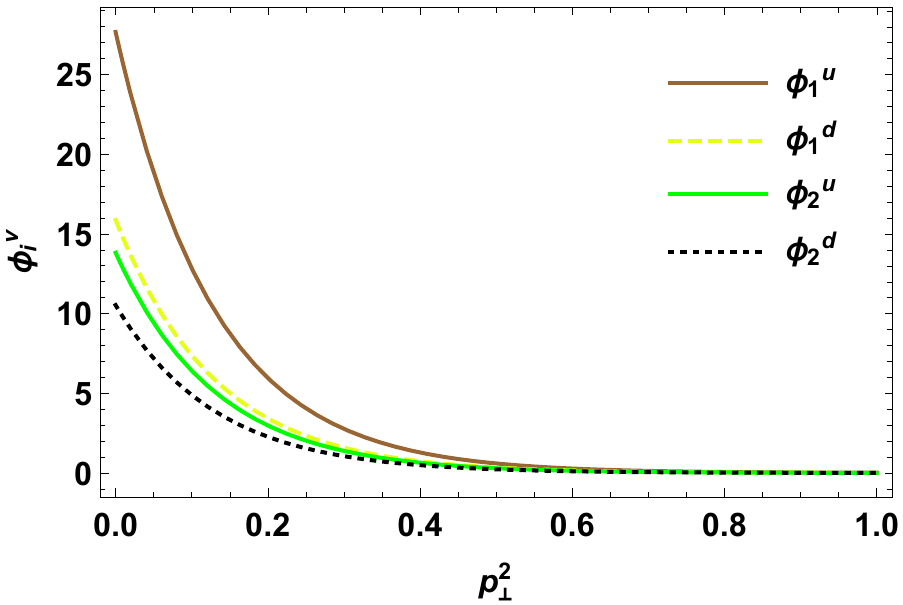}
\hspace{0.05cm}
\end{minipage}
\caption{\label{figphi} (Color online) (a) The wave functions $\varphi_i^\nu $ ($i=1,2; \nu = u, d)$ plotted with respect to  $x$ at ${\bf p_\perp^2}=0.2~\mathrm{GeV}^2$.
(b) The wave functions $\varphi_i^\nu $ plotted with respect to  ${\bf p_\perp^2}$ at $x=0.3$.}
\end{figure*}
\begin{figure*}
\centering
\begin{minipage}[c]{0.98\textwidth}
(a)\includegraphics[width=7.5cm,clip]{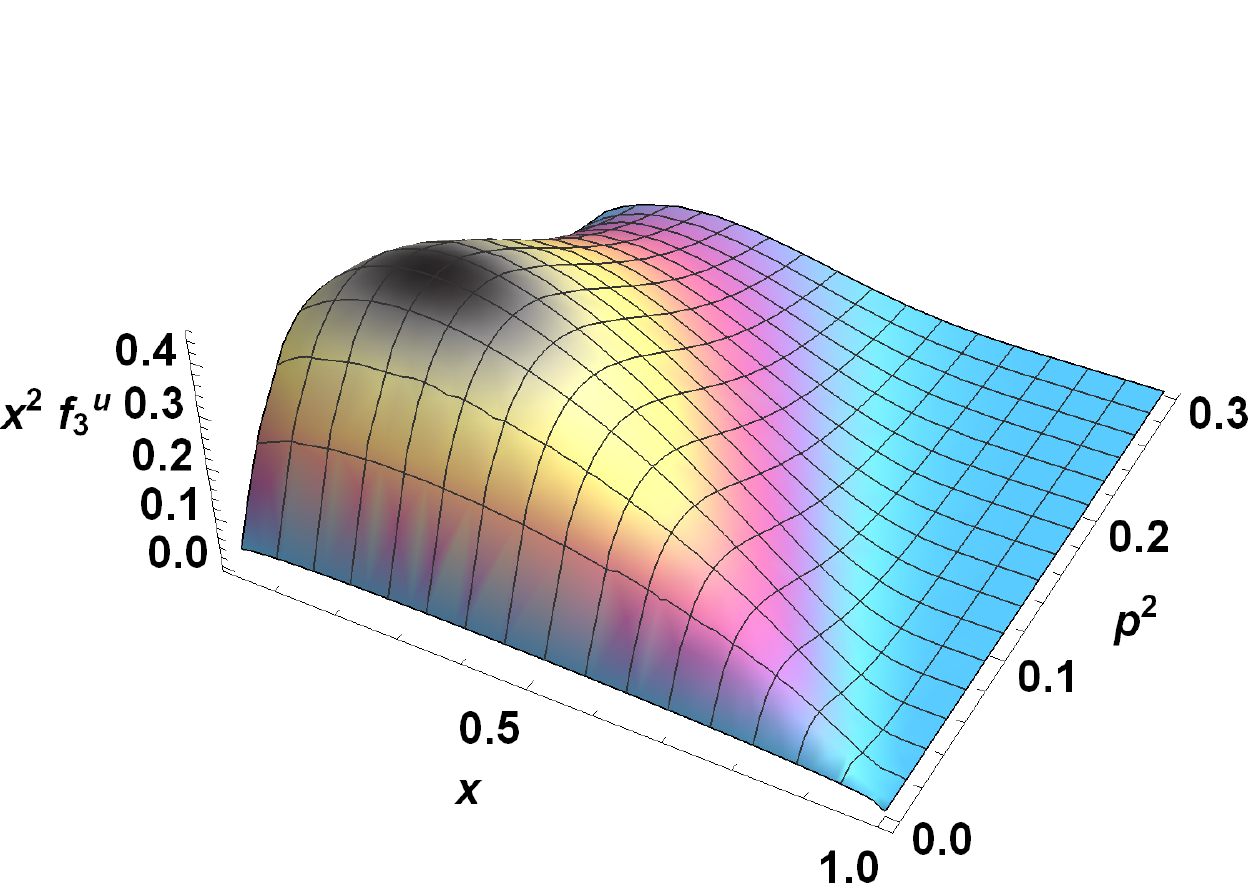}
\hspace{0.05cm}
(b)\includegraphics[width=7.5cm,clip]{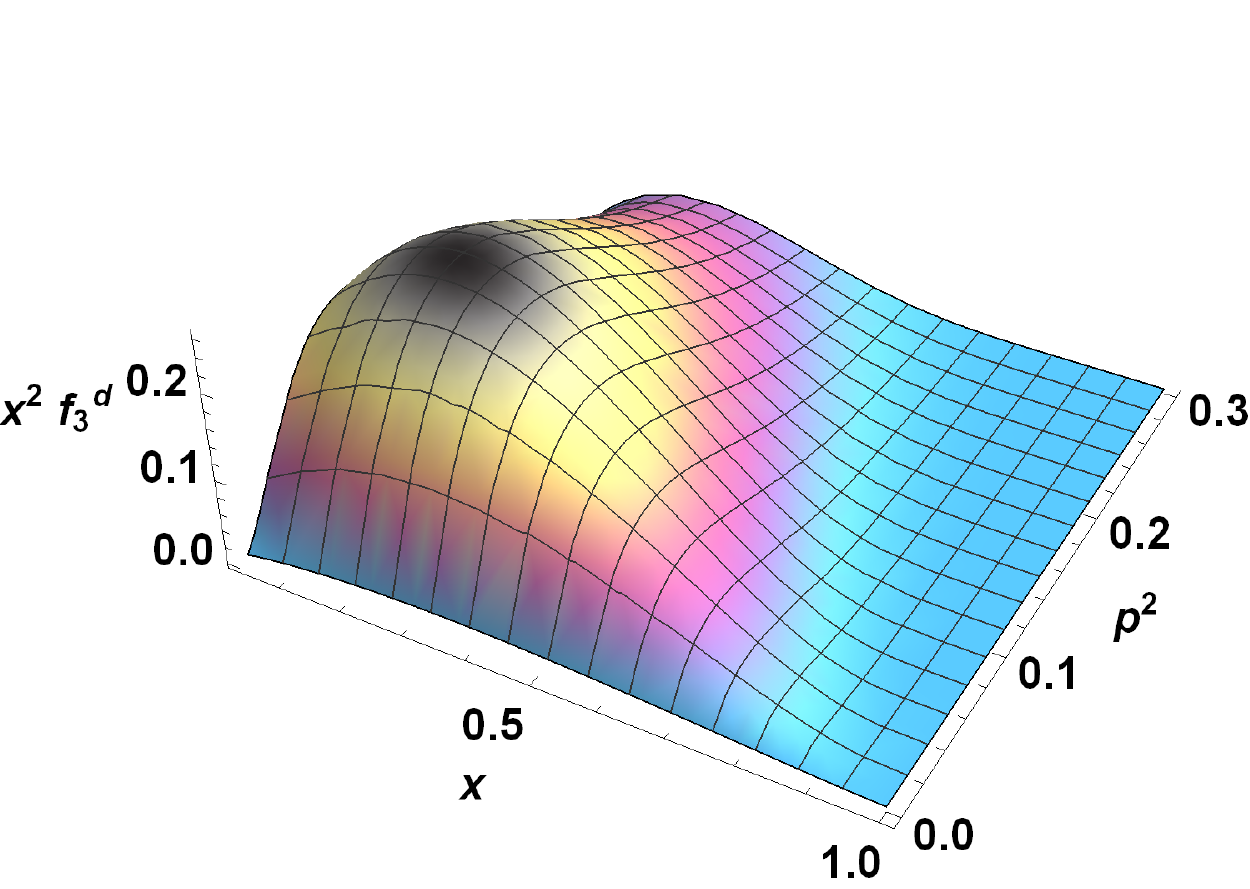}
\hspace{0.05cm}
\end{minipage}
\caption{\label{fig3d1} (Color online) The unpolarized TMD $x^2 f_{3}^{\nu}(x, {\bf p_\perp^2})$ plotted with respect to $x$ and ${\bf p_\perp^2}$. The left and right column correspond to $u$ and $d$ quarks sequentially.}
\end{figure*}
\begin{figure*}
\centering
\begin{minipage}[c]{0.98\textwidth}
(a)\includegraphics[width=7.5cm,clip]{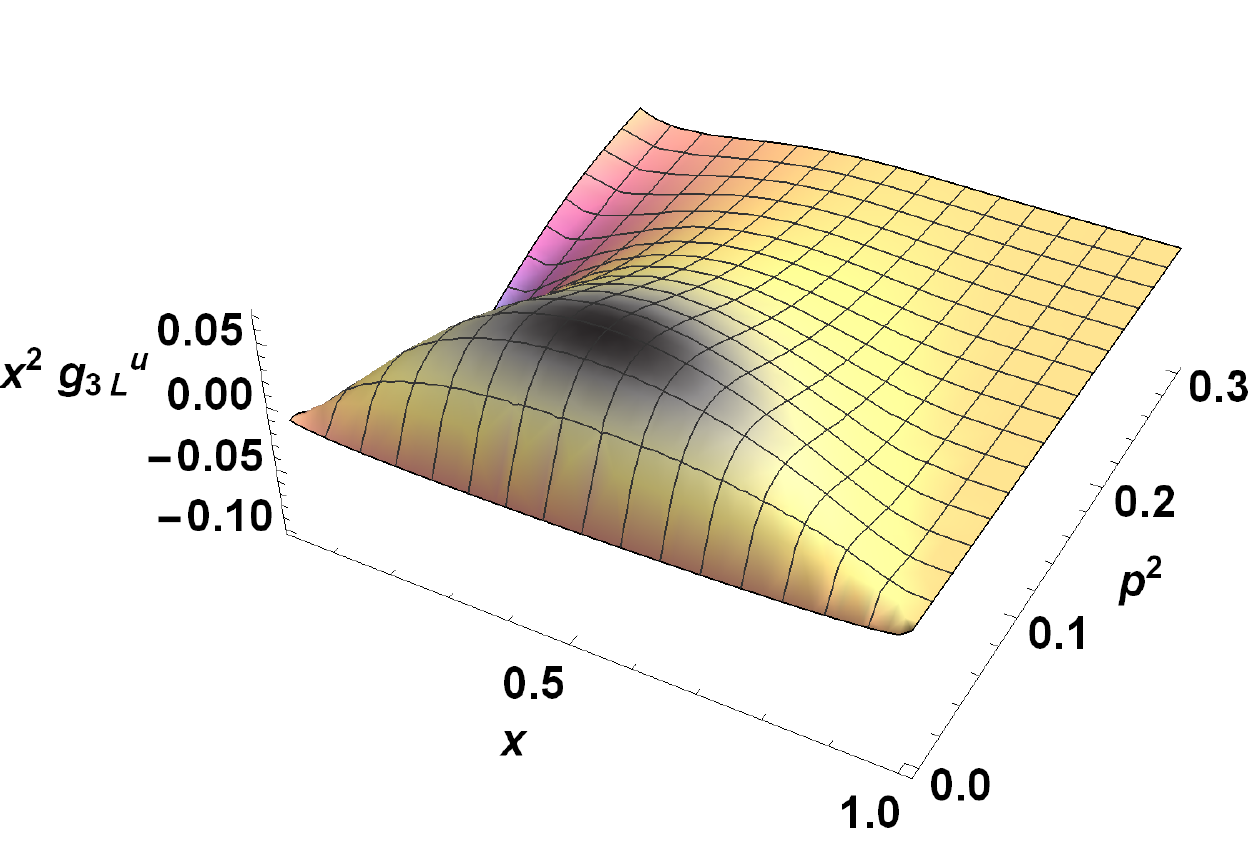}
\hspace{0.05cm}
(b)\includegraphics[width=7.5cm,clip]{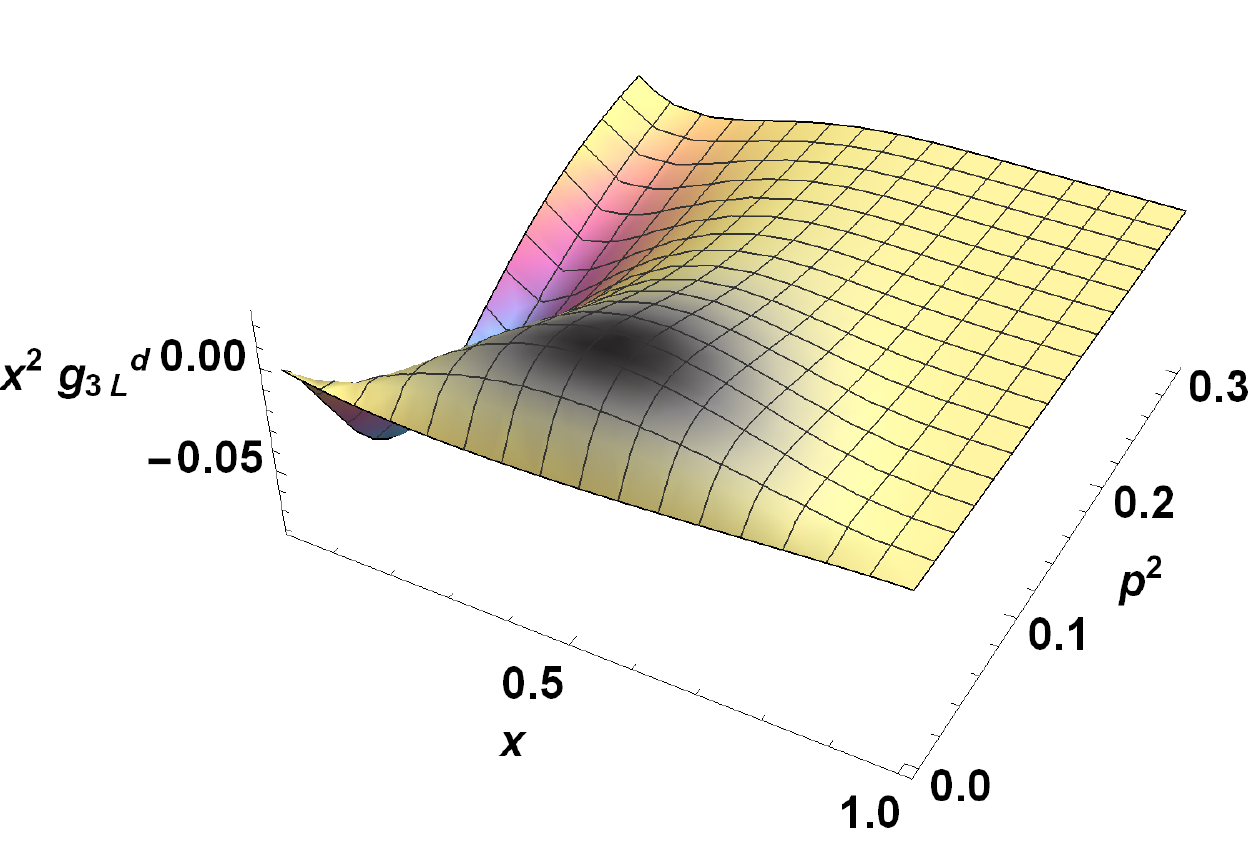}
\hspace{0.05cm}
(c)\includegraphics[width=7.5cm,clip]{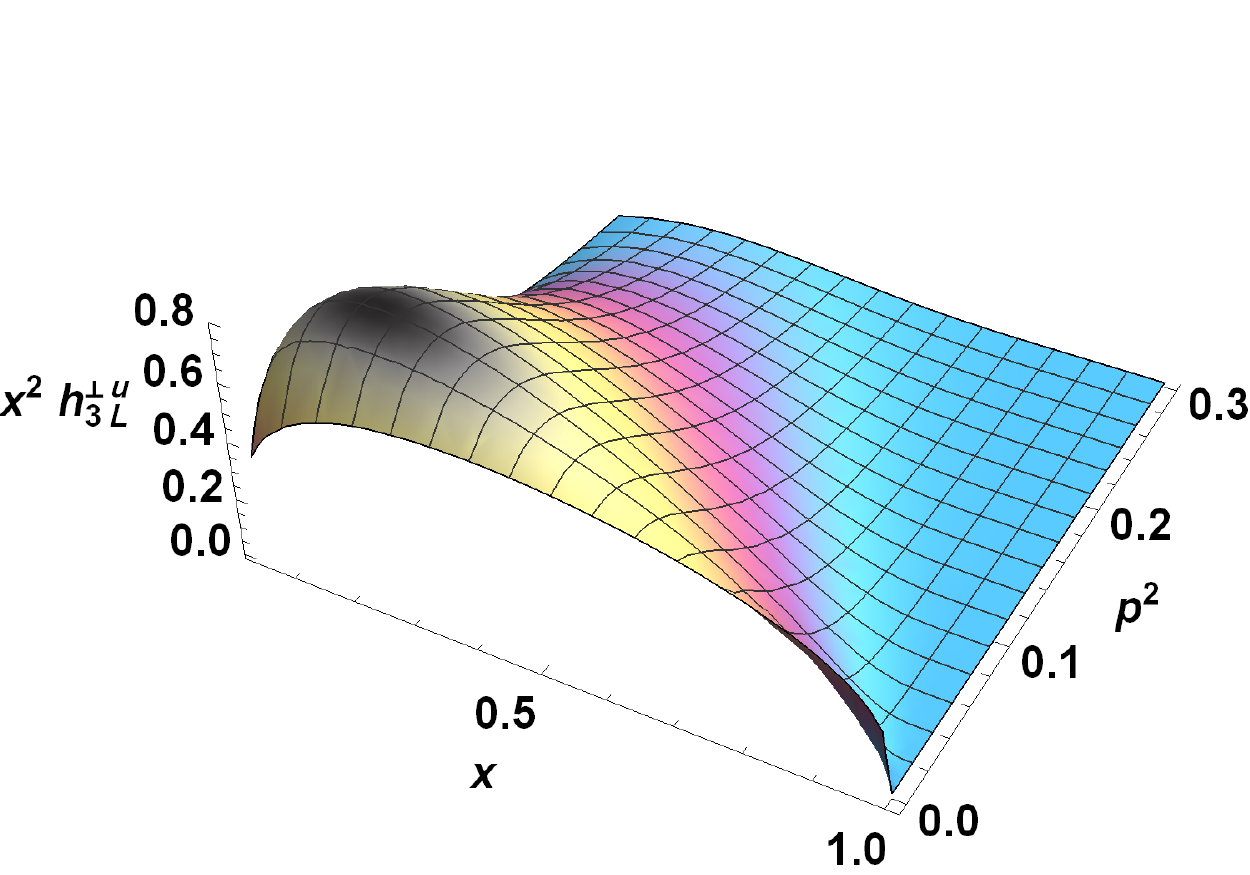}
\hspace{0.05cm}
(d)\includegraphics[width=7.5cm,clip]{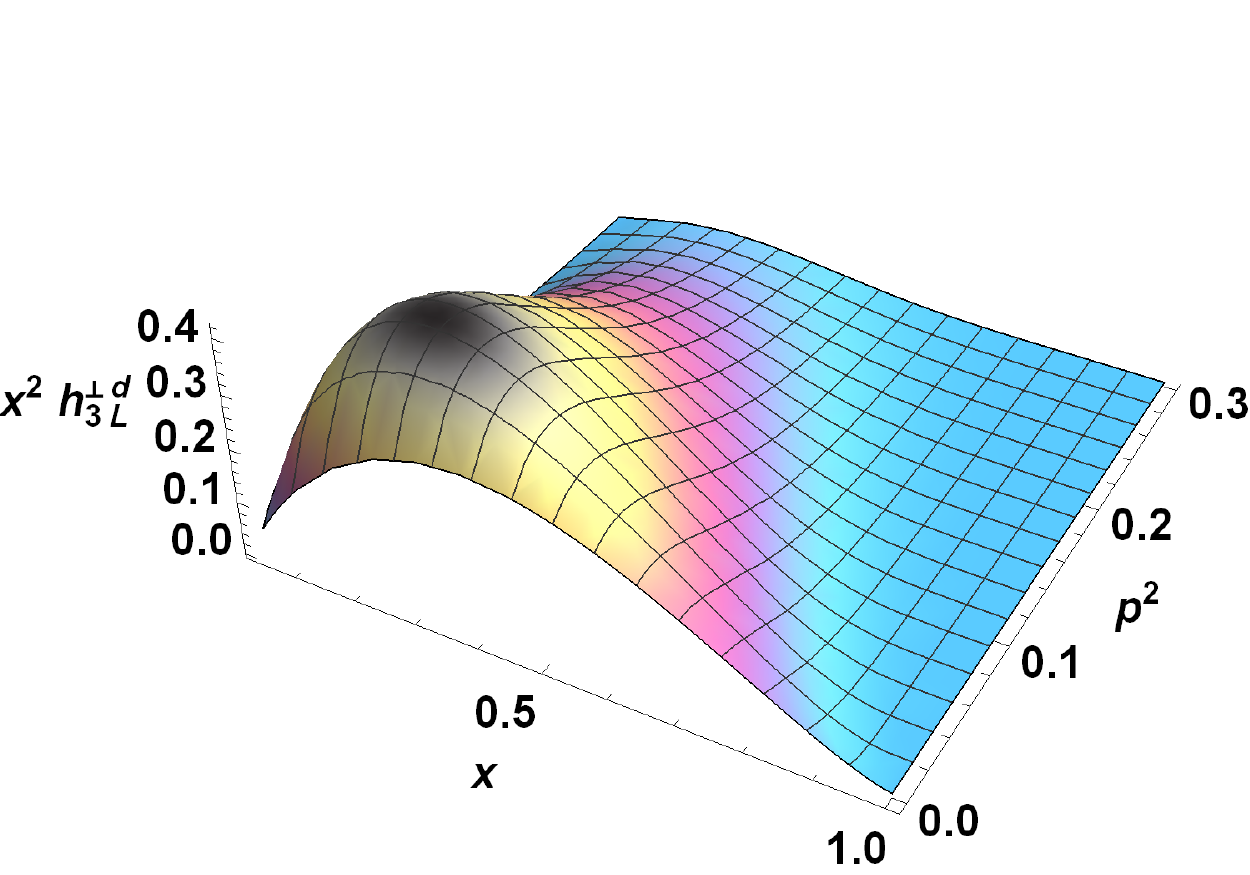}
\hspace{0.05cm}\\
\end{minipage}
\caption{\label{fig3d2} (Color online) The longitudinally polarized TMDs $x^2 g_{3L}^{\nu}(x, {\bf p_\perp^2})$ and $x^2 h_{3L}^{\perp\nu}(x, {\bf p_\perp^2})$ plotted with respect to $x$ and ${\bf p_\perp^2}$. The left and right column correspond to $u$ and $d$ quarks sequentially.}
\end{figure*}
\begin{figure*}
\centering
\begin{minipage}[c]{0.98\textwidth}
(a)\includegraphics[width=7.5cm]{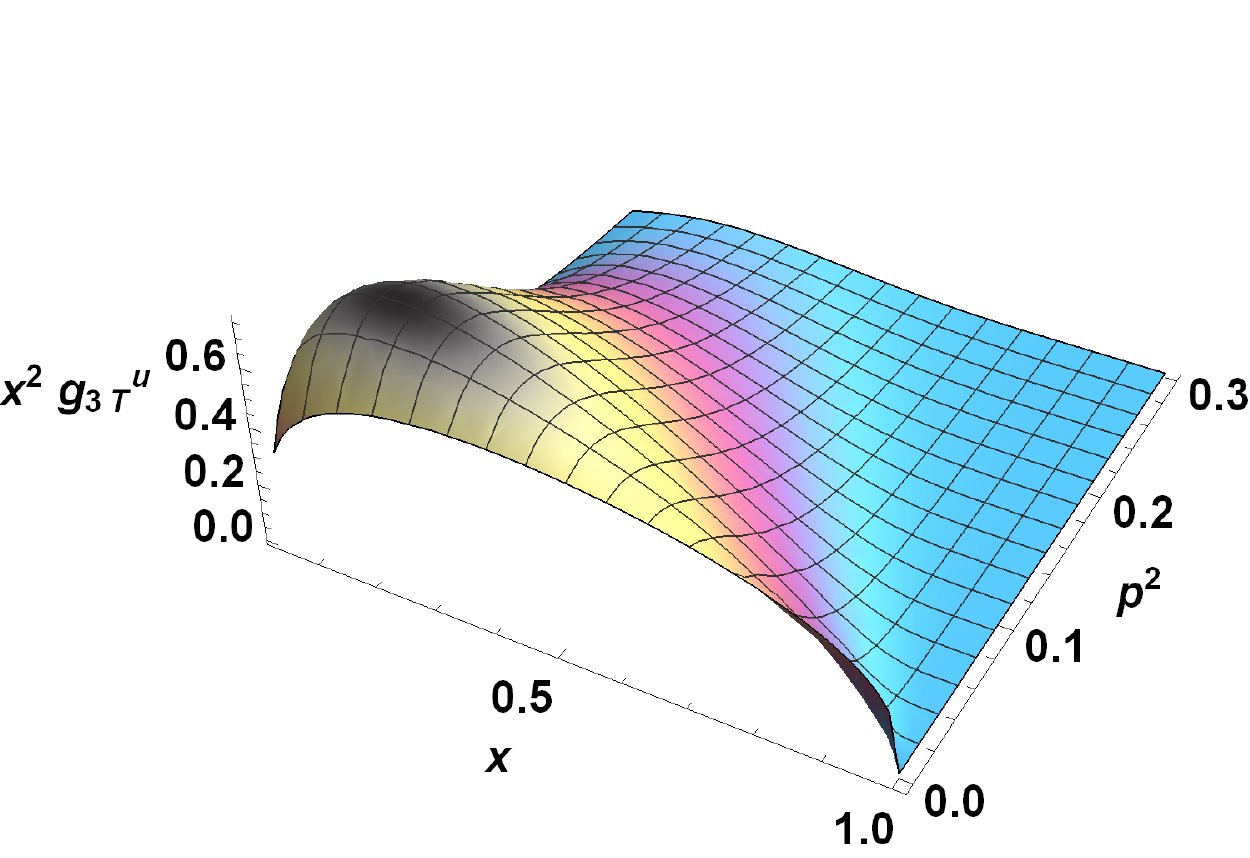}
\hspace{0.05cm}
(b)\includegraphics[width=7.5cm]{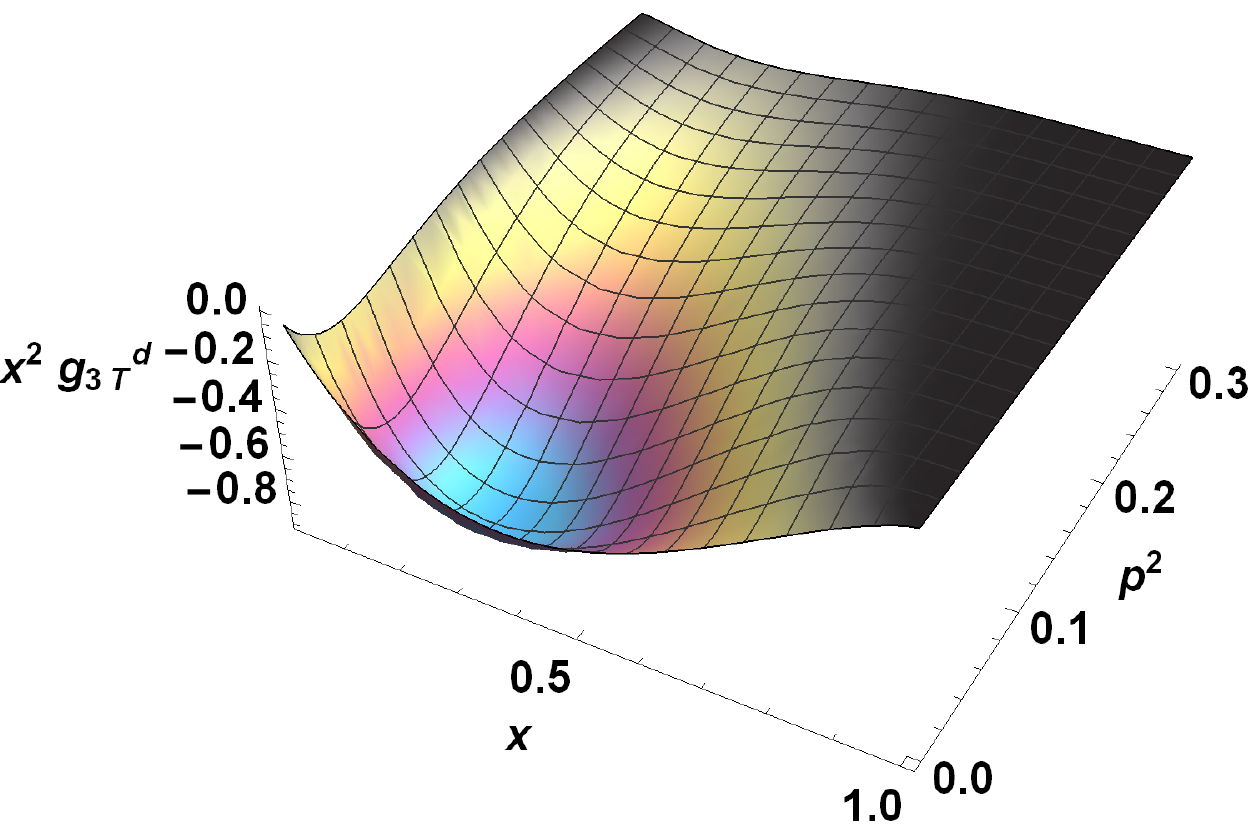}
\hspace{0.05cm}
(c)\includegraphics[width=7.5cm]{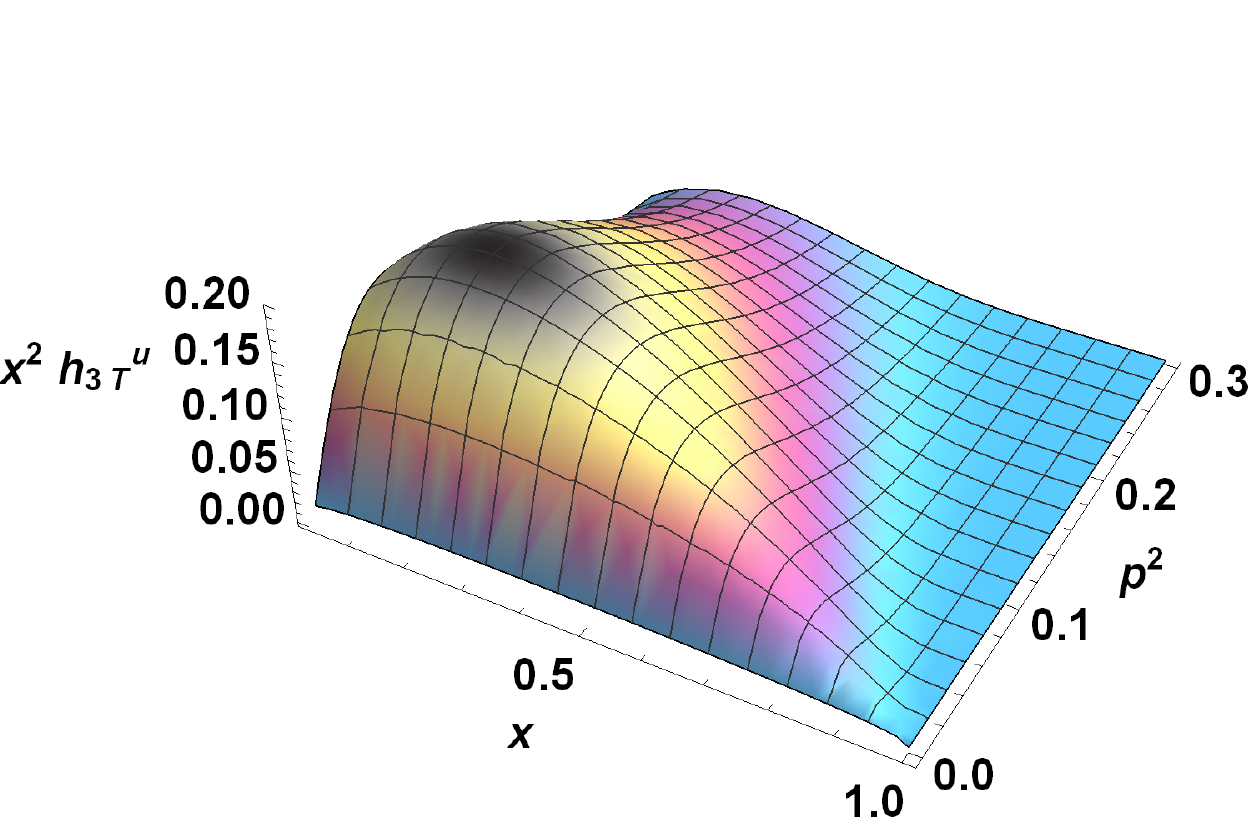}
\hspace{0.05cm}
(d)\includegraphics[width=7.5cm]{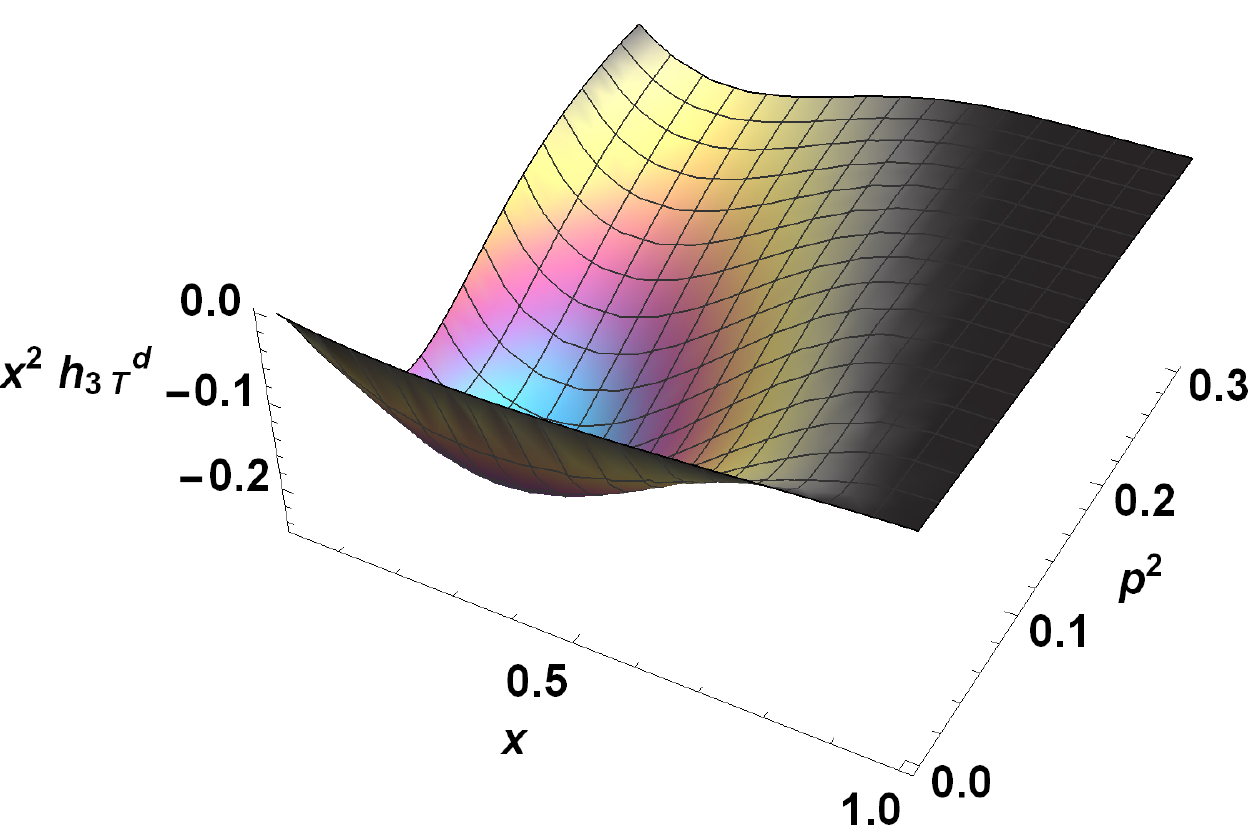}
\hspace{0.05cm}
(e)\includegraphics[width=7.5cm]{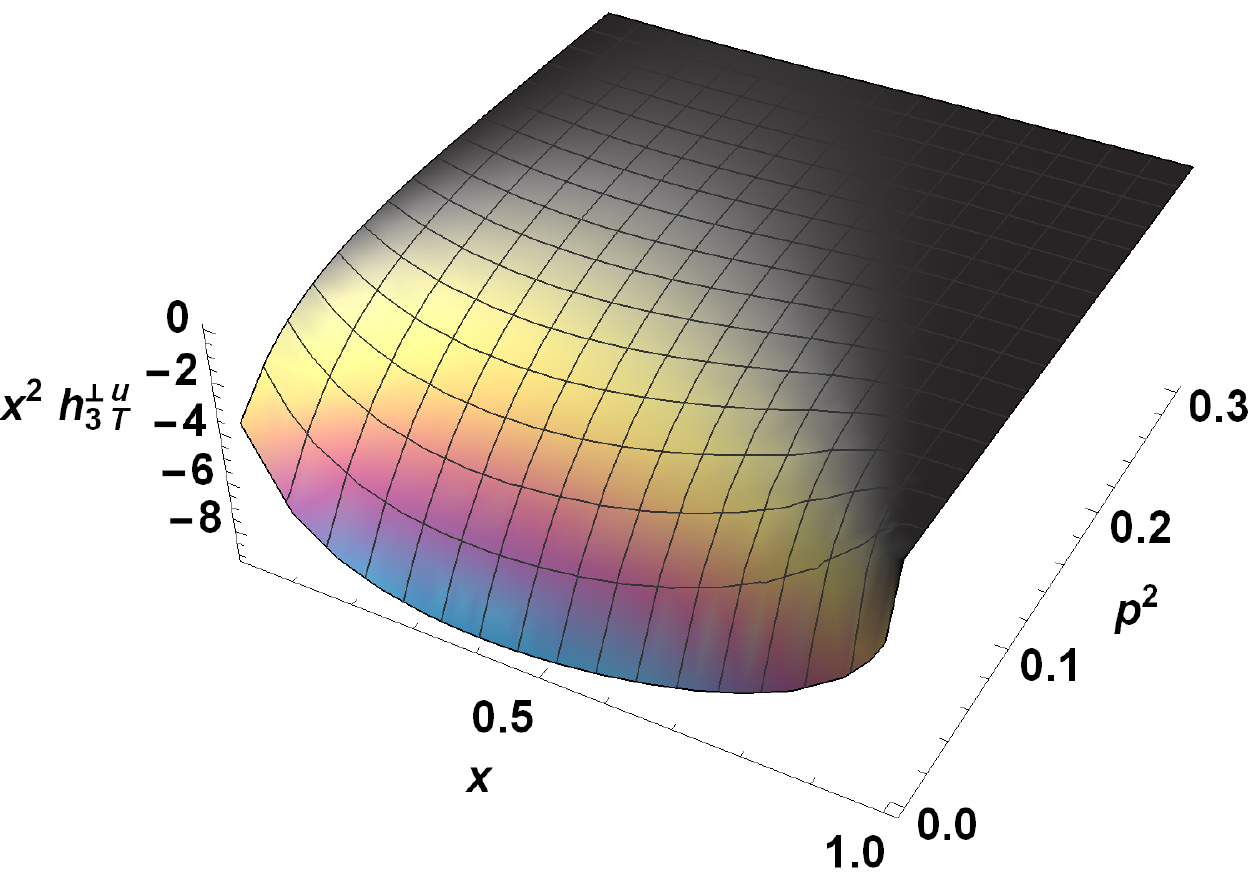}
\hspace{0.05cm}
(f)\includegraphics[width=7.5cm]{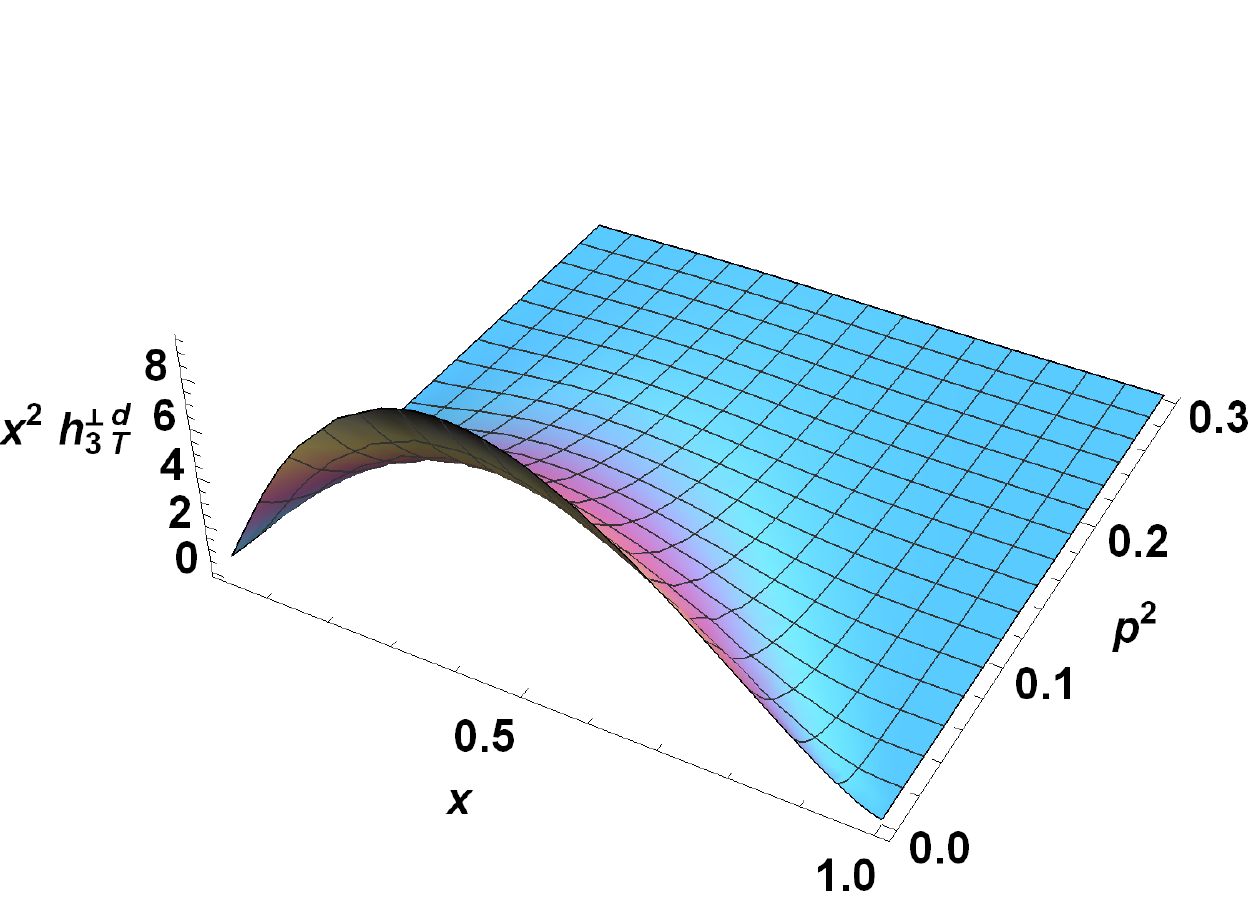}
\hspace{0.05cm}\\
\end{minipage}
\caption{\label{fig3d3} (Color online) The transversely polarized TMDs $x^2 {g}^{\nu  }_{3T}(x, {\bf p_\perp^2}),~x^2 {h}^{\nu  }_{3T}(x, {\bf p_\perp^2})$ and $x^2{h}^{\nu\perp}_{3T}(x, {\bf p_\perp^2})$ plotted with respect to $x$ and ${\bf p_\perp^2}$. The left and right column correspond to $u$ and $d$ quarks sequentially.}
\end{figure*}
\begin{figure*}
\centering
\begin{minipage}[c]{0.98\textwidth}
(a)\includegraphics[width=7.5cm]{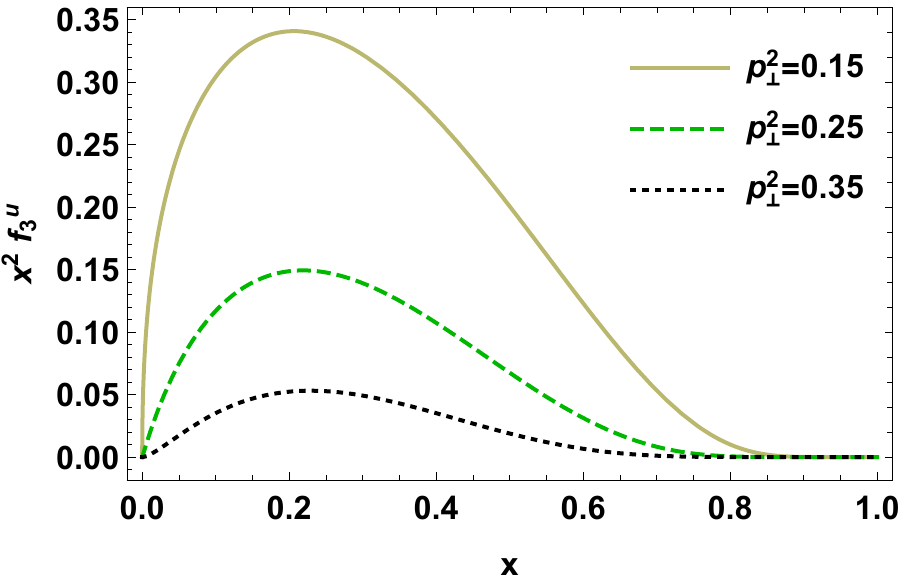}
\hspace{0.05cm}
(b)\includegraphics[width=7.5cm]{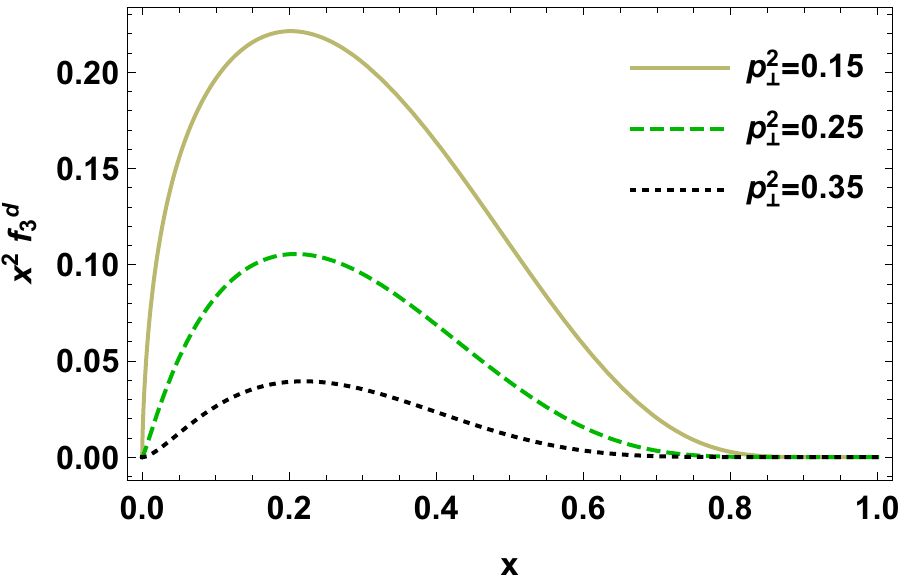}
\hspace{0.05cm}
\end{minipage}
\caption{\label{fig2dvx1} (Color online) The unpolarized TMD $x^2 f_{3}^{\nu}(x, {\bf p_\perp^2})$ plotted with respect to $x$ at different values of $ {\bf p_\perp^2}$, i.e., ${\bf p_\perp^2}=0.15~\mathrm{GeV}^2$ (olive green curve), ${\bf p_\perp^2}=0.25~\mathrm{GeV}^2$ (dashed green curve) and ${\bf p_\perp^2}=0.35~\mathrm{GeV}^2$ (dotted black curve). The left and right column correspond to $u$ and $d$ quarks sequentially.}
%
\end{figure*}
\begin{figure*}
\centering
\begin{minipage}[c]{0.98\textwidth}
(a)\includegraphics[width=7.5cm]{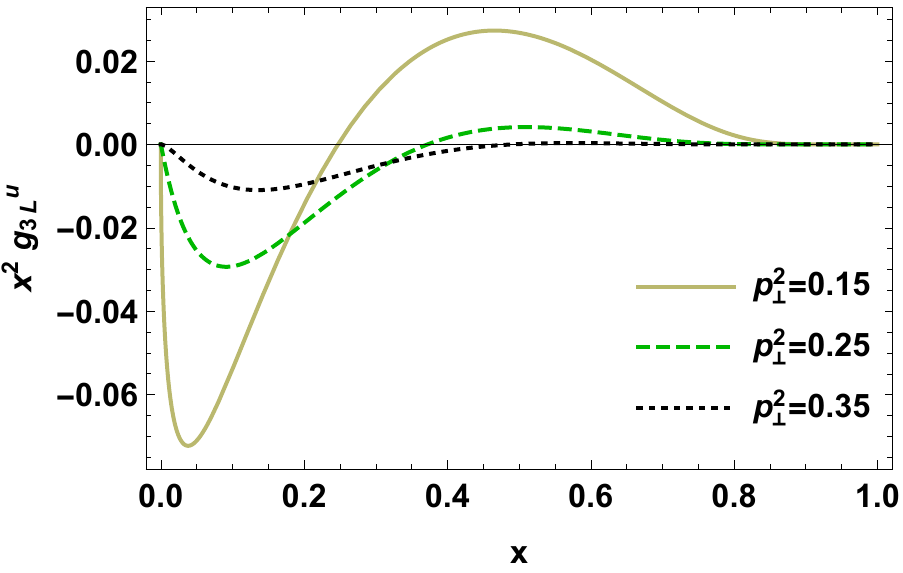}
\hspace{0.05cm}
(b)\includegraphics[width=7.5cm]{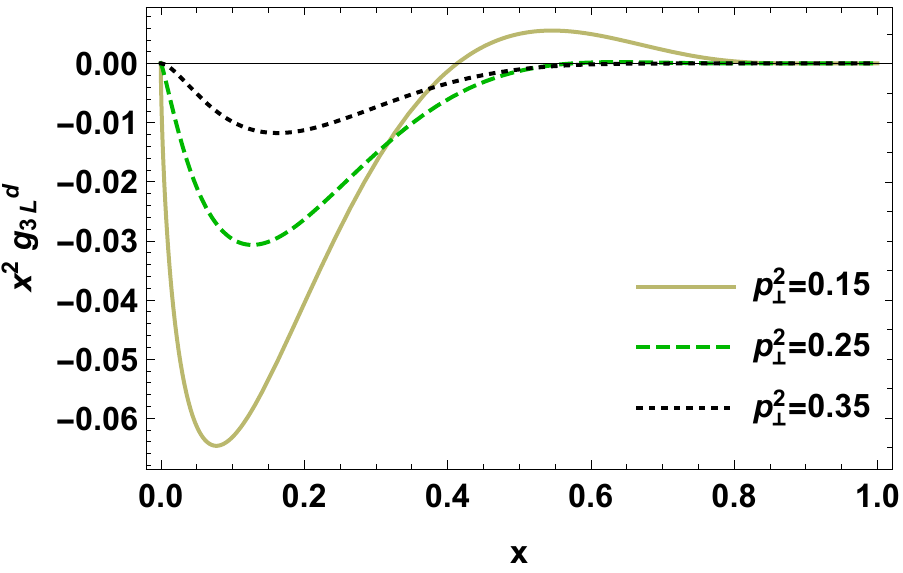}
\hspace{0.05cm}
(c)\includegraphics[width=7.5cm]{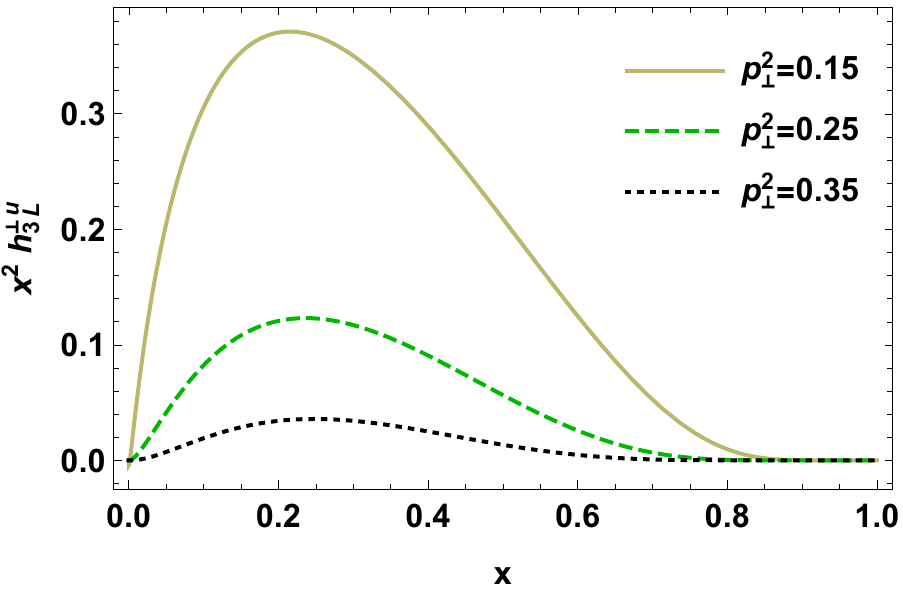}
\hspace{0.05cm}
(d)\includegraphics[width=7.5cm]{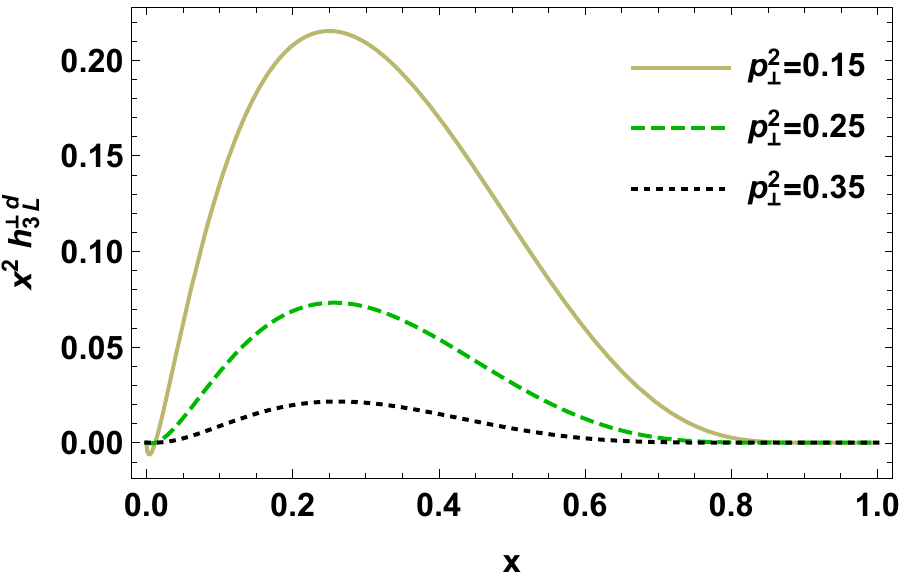}
\hspace{0.05cm}
\end{minipage}
\caption{\label{fig2dvx2} (Color online) The longitudinally polarized TMDs $x^2 g_{3L}^{\nu}(x, {\bf p_\perp^2})$ and $x^2 h_{3L}^{\perp\nu}(x, {\bf p_\perp^2})$ plotted with respect to $x$ at different values of $ {\bf p_\perp^2}$, i.e., ${\bf p_\perp^2}=0.15~\mathrm{GeV}^2$ (olive green curve), ${\bf p_\perp^2}=0.25~\mathrm{GeV}^2$ (dashed green curve) and ${\bf p_\perp^2}=0.35~\mathrm{GeV}^2$ (dotted black curve). The left and right column correspond to $u$ and $d$ quarks sequentially.}
%
\end{figure*}
\begin{figure*}
\centering
\begin{minipage}[c]{0.98\textwidth}
(a)\includegraphics[width=7.5cm]{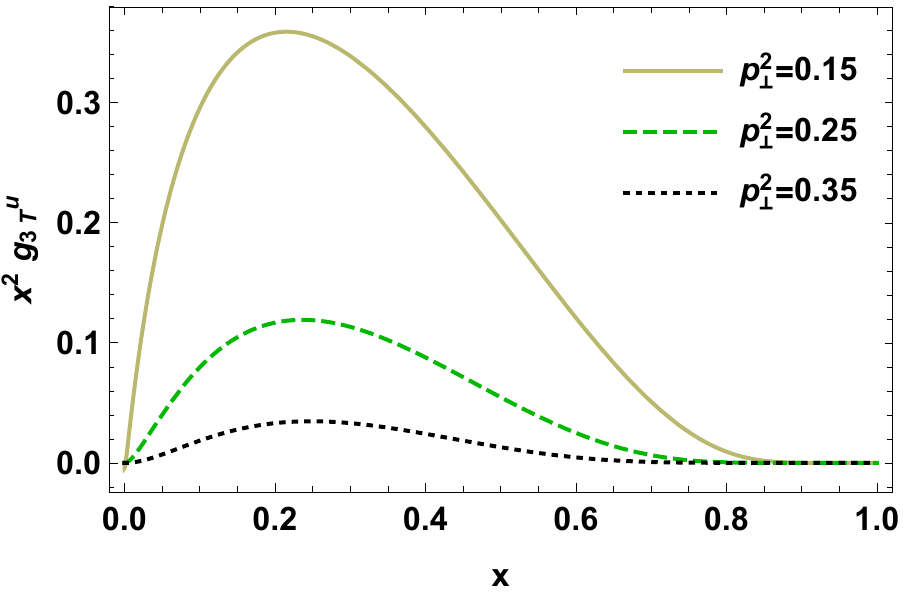}
\hspace{0.05cm}
(b)\includegraphics[width=7.5cm]{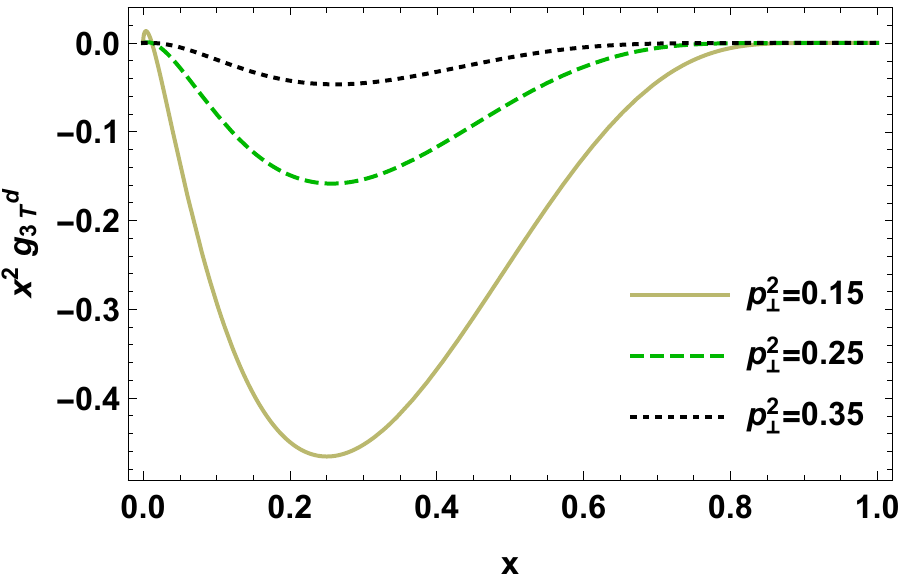}
\hspace{0.05cm}
(c)\includegraphics[width=7.5cm]{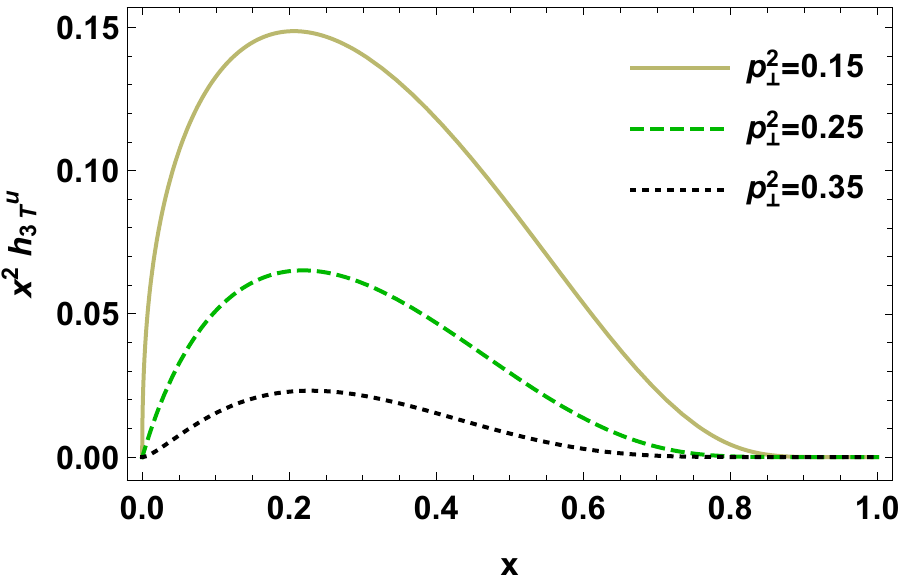}
\hspace{0.05cm}
(d)\includegraphics[width=7.5cm]{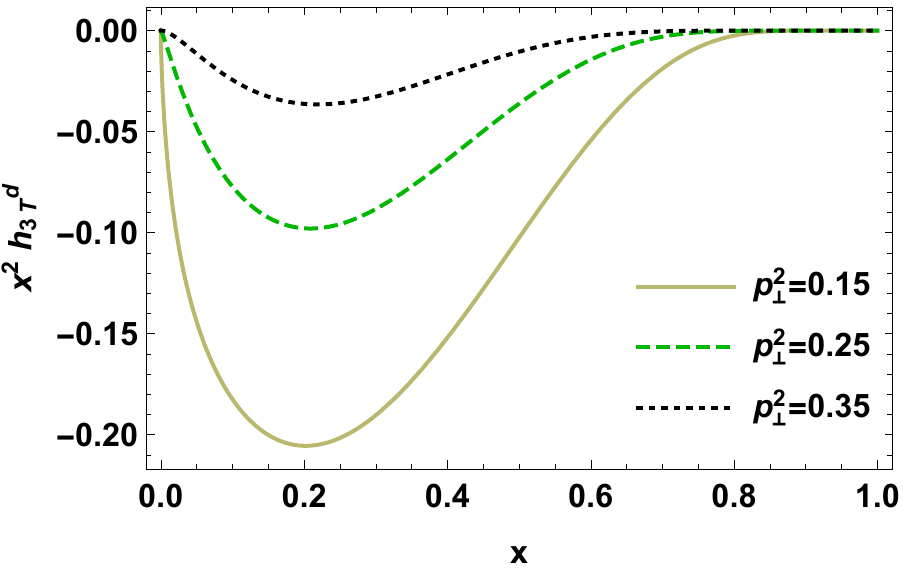}
\hspace{0.05cm}
(e)\includegraphics[width=7.5cm]{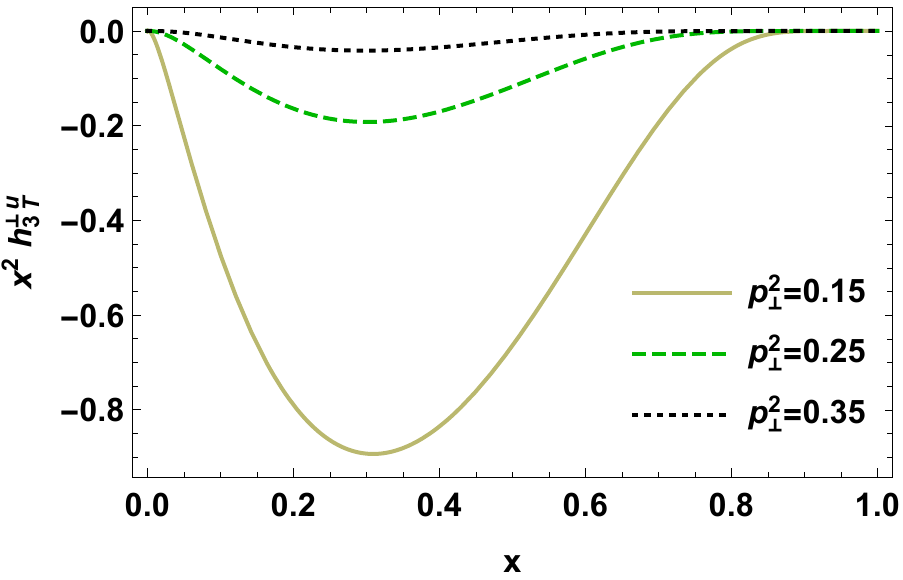}
\hspace{0.05cm}
(f)\includegraphics[width=7.5cm]{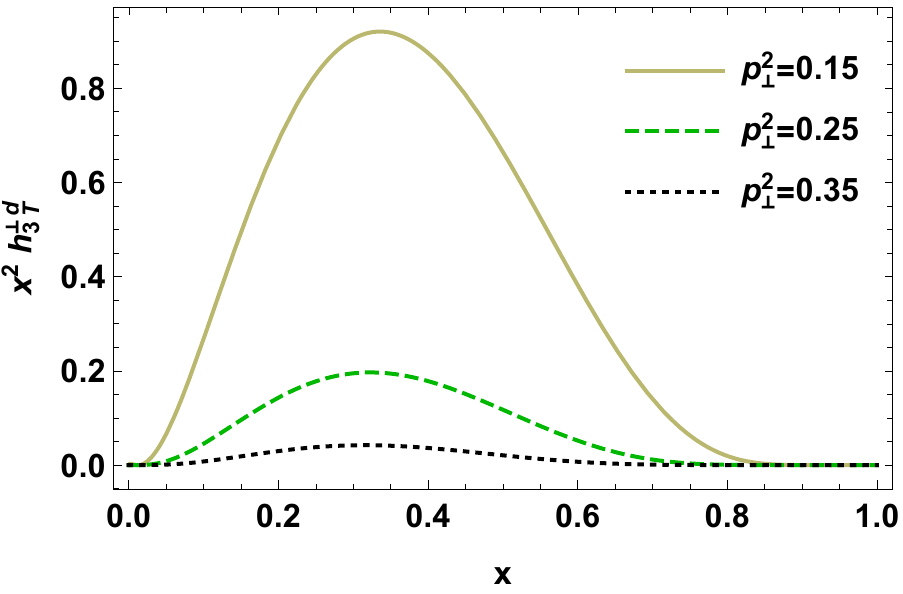}
\hspace{0.05cm}
\end{minipage}
\caption{\label{fig2dvx3} (Color online) The transversely polarized TMDs $x^2 {g}^{\nu  }_{3T}(x, {\bf p_\perp^2}),~x^2 {h}^{\nu  }_{3T}(x, {\bf p_\perp^2})$ and $x^2{h}^{\nu\perp}_{3T}(x, {\bf p_\perp^2})$ plotted with respect to $x$ at different values of $ {\bf p_\perp^2}$, i.e., ${\bf p_\perp^2}=0.15~\mathrm{GeV}^2$ (olive green curve), ${\bf p_\perp^2}=0.25~\mathrm{GeV}^2$ (dashed green curve) and ${\bf p_\perp^2}=0.35~\mathrm{GeV}^2$ (dotted black curve). The left and right column correspond to $u$ and $d$ quarks sequentially.}
%
\end{figure*}
\begin{figure*}
\centering
\begin{minipage}[c]{0.98\textwidth}
(a)\includegraphics[width=7.5cm]{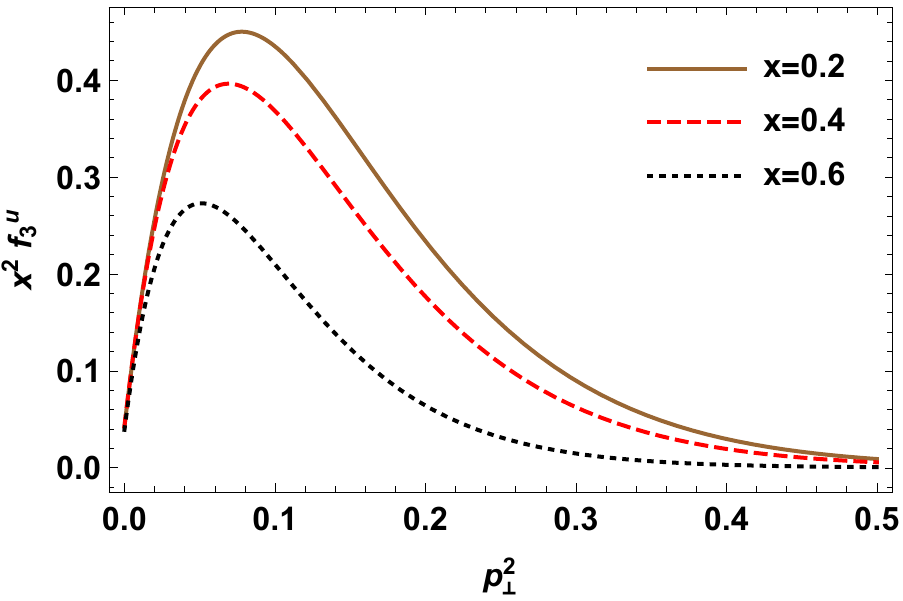}
\hspace{0.05cm}
(b)\includegraphics[width=7.5cm]{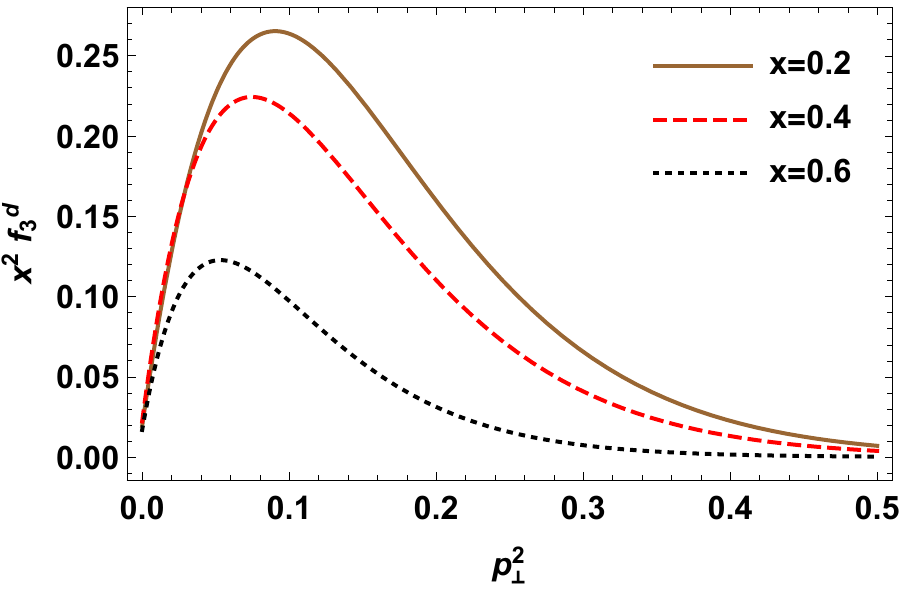}
\hspace{0.05cm}
\end{minipage}
\caption{\label{fig2dvp1} (Color online) The unpolarized TMD $x^2 f_{3}^{\nu}(x, {\bf p_\perp^2})$ plotted with respect to ${\bf p_\perp^2}$  at different values of $x$, i.e., $x=0.2$ (brown curve), $x=0.4$ (dashed red curve) and $x=0.6$ (dotted black curve). The left and right column correspond to $u$ and $d$ quarks sequentially.}
\end{figure*}
\subsection{Average Transverse Momentum $\langle\bfp^r(\Upsilon)\rangle^\nu$}\label{secavgtr}
The average transverse momenta ($ r=1 $) and the average square transverse momenta ($ r=2 $) for TMD $ {\Upsilon}^{\nu}(x,\bfp^2) $ in LFQDM is defined as
\be 
\langle\bfp^r(\Upsilon)\rangle^\nu= \frac{\int dx\int d^2p_\perp p^r_\perp {\Upsilon}^{\nu}(x,\bfp^2)}{\int dx\int d^2p_\perp {\Upsilon}^{\nu}(x,\bfp^2)}.
\label{Eqavg}
\ee 
\par The results of average transverse momenta and average transverse momenta squares for twist 4 T-even TMDs in LFQDM have been shown in Table \ref{tab_avgP4}. 
Since the parameterization of leading twist and twist-4 TMDs is similar, it is beneficial to compare our results with leading twist T-even TMD results. So, along with it we have tabulated average transverse momentum and average transverse momentum squares of leading twist T-even TMDs  from our model (LFQDM) in the Table \ref{tab_avgP4}. All results are in units of the respective value for $f_1^{\nu}$, which is $ \langle p_\perp \rangle^{u} =0.23~\mathrm{GeV}$, $\langle p_\perp \rangle^{d}=0.24~\mathrm{GeV}, \langle p_\perp^2 \rangle^{u} =0.066~\mathrm{GeV}^2$, $\langle p_\perp^2 \rangle^{d}=0.075~\mathrm{GeV}^2$. On a closer examination of Table \ref{tab_avgP4}, we observe that, similar to the leading twist TMD, $ f_{1}^{\nu}(x, {\bf p_\perp^2}) $ has the same average transverse momentum (/momentum square) value as $ h_{1T}^{\nu}(x, {\bf p_\perp^2}) $. Correspondingly their twist-4 partners $ f_{3}^{\nu}(x, {\bf p_\perp^2}) $ and $ h_{3T}^{\nu}(x, {\bf p_\perp^2}) $ also exhibit a similar trend.
Similarly, twist-4 TMDs $g_{3T}^{\nu}(x, {\bf p_\perp^2})$ and $h_{3L}^{\perp\nu}(x, {\bf p_\perp^2})$ have equal values, and in leading twist, values of TMD $g_{1T}^{\nu}(x, {\bf p_\perp^2})$ and $h_{1L}^{\perp\nu}(x, {\bf p_\perp^2})$ are equal.\par

Furthermore, the values of average transverse momentum and average square transverse momentum for TMD ${f}^{\nu  }_3(x, {\bf p_\perp^2})$ from LFCQM \cite{Lorce:2014hxa} have been compared with our findings in Table \ref{tab_avgP5}. The average transverse momentum has been written in units of $\mathrm{GeV}$ and average transverse momentum squares has been written in units of $\mathrm{GeV}^2$. It must be noted that these values are flavor dependent in our model whereas in LFCQM they are flavor independent. Values of $ \langle p_\perp \rangle^{u}$ and $ \langle p_\perp \rangle^{d}$ are similar in both models but $\langle p_\perp^2 \rangle^{u}$ and $\langle p_\perp^2 \rangle^{d}$ values in our model are slightly smaller when compared with LFCQM \cite{Lorce:2014hxa}.
\begin{table}[h]
	\centering
	\begin{tabular}{||c||c|c|c|c|c|c||c|c|c|c|c|c||}
		\hline
		\hline
		\text{TMD} $\Upsilon  $~~&~~$ f_{3}^{\nu} $~~&~~$  g_{3L}^{\nu} $~~&~~$ g_{3T}^{\nu} $~~&~~$ h_{3L}^{\perp\nu} $~~&~~$ h_{3T}^{\nu} $~~&~~$h_{3T}^{\perp \nu}$~~&~~$ f_{1}^{\nu} $~~&~~$  g_{1L}^{\nu} $~~&~~$ g_{1T}^{\nu} $~~&~~$ h_{1L}^{\perp\nu} $~~&~~$ h_{1T}^{\nu} $~~&~~$h_{1T}^{\perp \nu}$\\
		\hline
		\hline
		$ \langle p_\perp \rangle^{u} $~~&~~$ 1.14 $~~&~~$ 1.13 $~~&~~$ 4.02 $~~&~~$ 4.02 $~~&~~$ 1.14 $~~&~~$ 0.76$~~&~~$ 1.00 $~~&~~$ 0.78 $~~&~~$ 0.92 $~~&~~$ 0.92 $~~&~~$ 1.00 $~~&~~$ 0.90$ \\
		\hline
		$ \langle p_\perp \rangle^{d} $~~&~~$ 1.09 $~~&~~$ 1.10 $~~&~~$ 0.80 $~~&~~$ 0.80 $~~&~~$ 1.09 $~~&~~$ 0.58$~~&~~$ 1.00 $~~&~~$ 0.54 $~~&~~$ 0.87 $~~&~~$ 0.87 $~~&~~$ 1.00 $~~&~~$ 0.85$ \\
		\hline
		$ \langle p_\perp^2 \rangle^{u} $~~&~~$ 1.11 $~~&~~$ 1.10 $~~&~~$ 6.50 $~~&~~$ 6.50 $~~&~~$ 1.11 $~~&~~$ 0.59  $~~&~~$ 1.00 $~~&~~$ 0.59 $~~&~~$ 0.85 $~~&~~$ 0.85 $~~&~~$ 1.00 $~~&~~$ 0.81$\\
		\hline
		$ \langle p_\perp^2 \rangle^{d} $~~&~~$ 1.04 $~~&~~$ 1.06 $~~&~~$ 0.57 $~~&~~$ 0.57 $~~&~~$ 1.04 $~~&~~$ 0.35 $~~&~~$ 1.00 $~~&~~$ 0.19 $~~&~~$ 0.78 $~~&~~$ 0.78 $~~&~~$ 1.00 $~~&~~$ 0.74 $\\
		\hline
		\hline
	\end{tabular}
	\caption{Average transverse momentum and average transverse momentum squares for twist-4 and leading twist T-even TMDs in LFQDM.}
	\label{tab_avgP4} 
\end{table}

\begin{table}[h]
	\centering
	\begin{tabular}{||c||c|c||}
		\hline
		\hline
		$ \text{MODEL} $~~&~~$ f_{3}^{\nu}~\text{(LFQDM)} $~~&~~$ f_{3}^{\nu}~\text{(LFCQM)} $\\
		\hline
		\hline
		$ \langle p_\perp \rangle^{u} $~~&~~$ 0.26 $~~&~~$ 0.28 $ \\
		$ \langle p_\perp \rangle^{d} $~~&~~$ 0.27 $~~&~~$ 0.28 $\\
		$ \langle p_\perp^2 \rangle^{u} $~~&~~$ 0.07 $~~&~~$ 0.11 $\\
		$ \langle p_\perp^2 \rangle^{d} $~~&~~$ 0.08 $~~&~~$ 0.11 $\\
		\hline
		\hline
	\end{tabular}
	\caption{Comparison of average transverse momentum in units of $\mathrm{GeV}$ and average transverse momentum squares in units of $\mathrm{GeV}^2$ for TMD ${f}^{\nu  }_3(x, {\bf p_\perp^2})$ in LFQDM (our model) and LFCQM \cite{Lorce:2014hxa}.}
	\label{tab_avgP5} 
\end{table}
\subsection{$x-p_\perp^2$ dependence}\label{xdep}
Before we can comprehend the behaviour of TMDs, we must realize the nature of the wave functions $\varphi_i^\nu $, on which it depends via  Eqs. {\eqref{eef3s}-\eqref{eeh3tpv}}. In Fig. \ref{figphi} (a), the wave functions  $\varphi_i^\nu $ are illustrated with respect to $x$ at a fixed value of ${\bf p_\perp^2}=0.2~\mathrm{GeV}^2 $ to explore their dependency on longitudinal momentum fraction $x$. The value of ${\bf p_\perp^2}$ is assumed to be $0.2$ in order to clearly project the variation of the wave functions. At larger ${\bf p_\perp^2}$ levels, the wavefunctions pertaining to $u$ and $d$ quarks, as well as $\varphi_1^u $, $\varphi_1^d $, $\varphi_2^u $ and $\varphi_2^d $ do not differ significantly. In Fig. \ref{figphi} (b), wave functions $\varphi_i^\nu $ are displayed with respect to ${\bf p_\perp^2}$ at a fixed $x=0.3$ for ${\bf p_\perp^2}$ dependence study. Notably, as the wave functions in Fig. \ref{figphi} (a), peak around $x=0.3$, we have fixed the value of $x$ to $0.3$ so that the dependence of $\varphi_i^\nu $ may be projected out clearly. In Fig. \ref{figphi} (a) and \ref{figphi} (b), the plots for $\varphi_1^u $, $\varphi_1^d $, $\varphi_2^u $ and $\varphi_2^d $ are represented by brown, dashed yellow, green, and dotted black curves, respectively. It is important to note that for ${\bf p_\perp^2}$ values larger than or equal to $0.5~\mathrm{GeV}^2$, the value of the wave function $\varphi_i^\nu$ is $0$. Even at ${\bf p_\perp^2}=0.3~\mathrm{GeV}^2$, its magnitude is negligibly small. Keeping this in mind, we have shown our future graphs for ${\bf p_\perp^2}$ values up to $0.3~\mathrm{GeV}^2$.\par
In order to understand the dependence of twist-4 T-even TMDs with simultaneous change in variables $x$ and $\bfp^2$, we have plotted their 3-D variation. Unpolarized \big($x^2~{f}^{\nu  }_3(x, {\bf p_\perp^2})$\big), longitudinally polarized \big($x^2~{g}^{\nu  }_{3L}(x, {\bf p_\perp^2})$ and $x^2~{h}^{\nu\perp}_{3L}(x, {\bf p_\perp^2})$\big) and transversely polarized \big( ${g}^{\nu  }_{3T}(x, {\bf p_\perp^2}),~{h}^{\nu  }_{3T}(x, {\bf p_\perp^2})$ and 
${h}^{\nu\perp}_{3T}(x, {\bf p_\perp^2})$\big) TMDs are plotted with variables $x$ and $\bfp^2$ for up quark (left column) and down quark (right column) in Fig. (\ref{fig3d1}), (\ref{fig3d2}) and (\ref{fig3d3}) respectively. Firstly, we start our discussion with the unpolarized TMDs function $x^2~{f}^{\nu  }_3(x, \bfp^2)$ which is positive for all values of $x$ and $\bfp^2$  for both $u$ and $d$ quarks as represented by Fig. \ref{fig3d1} (a) and \ref{fig3d1} (b) respectively. It is closely related to its leading twist unpolarized partner ${f}^{\nu  }_1(x, {\bf p_\perp^2})$ via Eq. (\ref{cef3}). In Fig. \ref{fig3d2} (a) and Fig. \ref{fig3d2} (b), the longitudinally polarized function $x^2~{g}^{\nu  }_{3L}(x, {\bf p_\perp^2})$ is plotted for $u$ and $d$ quarks in left and right column sequentially. Although for $d$ quarks it is not visible in 3-D plot but it possess both positive and negative values with the variation in values of $x$ and $\bfp^2$. Unlike the unpolarized one, not only it has the contribution from leading twist TMD ${g}^{\nu  }_{1L}(x, {\bf p_\perp^2})$ but also from the ${h}^{\nu\perp}_{1L}(x, {\bf p_\perp^2})$ as shown in Eq. (\ref{ceg3l}).
Similarly, Eq. (\ref{ceh3lp}) shows that ${h}^{\nu\perp}_{3L}(x, {\bf p_\perp^2})$ is a mixture of both the longitudinally polarized leading twist TMDs function ${h}^{\nu\perp}_{1L}(x, {\bf p_\perp^2})$ and ${g}^{\nu  }_{1L}(x, {\bf p_\perp^2})$ like ${g}^{\nu  }_{3L}(x, {\bf p_\perp^2})$. But, contrary to plots of $x^2~{g}^{\nu  }_{3L}(x, {\bf p_\perp^2})$, $x^2~{h}^{\nu\perp}_{3L}(x, {\bf p_\perp^2})$ has a range of only positive values for both $u$ and $d$ quarks in $x$-$\bfp^2$ domain as plotted in Fig. \ref{fig3d2} (c) and Fig. \ref{fig3d2} (d). This may be due to the different dominance for positive or negative values of both the terms at different values of $x$ and $\bfp^2$. The transversely polarized TMD ${g}^{\nu  }_{3T}(x, {\bf p_\perp^2})$ is positive (negative) for all values of $x$ and $\bfp^2$ for $u$ ($d$) quarks as represented by Fig. \ref{fig3d3} (a) \big(\ref{fig3d3} (b)\big) . It is related to leading twist T-even TMDs via Eq. (\ref{ceg3t}). In Fig. \ref{fig3d3} (c) \big(\ref{fig3d3} (d)\big) ${h}^{\nu  }_{3T}(x, {\bf p_\perp^2})$ is plotted which is  positive (negative) for all values of $x$ and $\bfp^2$ for $u$ ($d$) quarks and it is related to its leading twist partner ${h}^{\nu  }_{1T}(x, {\bf p_\perp^2})$ via Eq. (\ref{ceh3t}). In Eq. (\ref{ceh3tp}), we have shown the relation of transversely polarized TMD ${h}^{\nu\perp}_{3T}(x, {\bf p_\perp^2})$ with leading twist TMDs. This TMD is negative (positive) for all values of $x$ and $\bfp^2$ for $u$ ($d$) quarks as represented by Fig. \ref{fig3d3} (e) \big(\ref{fig3d3} (f)\big). 
\par
To have a closer look on TMDs for its dependence on the longitudinal momentum fraction $x$, we have plotted the 2-dimensional (2-D) variation of TMDs with respect to $x$. In Fig. (\ref{fig2dvx1}), (\ref{fig2dvx2}) and (\ref{fig2dvx3}), the TMDs are plotted with respect to $x$ at different values of $ {\bf p_\perp^2}$, i.e., ${\bf p_\perp^2}=0.15~\mathrm{GeV}^2$ (olive green curve), ${\bf p_\perp^2}=0.25~\mathrm{GeV}^2$ (dashed green curve) and ${\bf p_\perp^2}=0.35~\mathrm{GeV}^2$ (dotted black curve). The left and right column correspond to $u$ and $d$ quarks respectively. It is found that, within any plot, as the chosen value of $ {\bf p_\perp^2}$ increases, the amplitude of TMDs decreases. For unpolarized and longitudinally polarized TMDs, no flip in the sign takes place in the shape of plot while changing the flavor from $u$ to $d$ quarks or vice versa, whereas for transversely polarized TMDs a flip in sign is observed. In Fig. \ref{fig2dvx1} (a) and \ref{fig2dvx1} (b) the TMD $x^2 f_{3}^{\nu}(x, {\bf p_\perp^2})$ first increases with the increase in longitudinal momentum fraction $x$, reaches a maxima and then decreases. This is true for both $u$ and $d$ quarks. The TMD $x^2 g_{3L}^{\nu}(x, {\bf p_\perp^2})$, with the increase in longitudinal momentum fraction, shows both minima and maxima with a node in between them as shown in Fig. \ref{fig2dvx2} (a) and \ref{fig2dvx2} (b). This suggests that the distribution flips sign at a particular value of longitudinal momentum fraction carried by the quark. In Fig. \ref{fig2dvx2} (c) and \ref{fig2dvx2} (d), with the variation in longitudinal momentum fraction $x$, the TMD $~x^2 h_{3L}^{\perp\nu}(x, {\bf p_\perp^2})$ surprisingly shows a trend similar to $x^2 f_{3}^{\nu}(x, {\bf p_\perp^2})$, although their explicit expressions are different. Ignoring the behaviour at the value of longitudinal momentum fraction close to $0$, we have observed that with increase in $x$ the TMD ${g}^{\nu  }_{3T}(x, {\bf p_\perp^2})$ first increases (decreases) and then decreases (increases) showing a maxima (minima) as shown in Fig. \ref{fig2dvx3} (a) \big(\ref{fig2dvx3} (b)\big). We have plotted the transversely polarized TMD ${h}^{\nu  }_{3T}(x, {\bf p_\perp^2})$ in Fig. \ref{fig2dvx3} (c) \big(\ref{fig2dvx3} (d)\big), which with the change in the longitudinal momentum fraction $x$, shows a trend similar to ${g}^{\nu  }_{3T}(x, {\bf p_\perp^2})$ except that there are no fluctuations here in low $x$ regime. In Fig. \ref{fig2dvx3} (e) \big(\ref{fig2dvx3} (f)\big), the transversely polarized TMD ${h}^{\nu\perp}_{3T}(x, {\bf p_\perp^2})$ is plotted which with increase in the longitudinal momentum fraction $x$, first decreases (increases) and then increases (decreases) showing a minima (maxima).
\begin{figure*}
\centering
\begin{minipage}[c]{0.98\textwidth}
(a)\includegraphics[width=7.5cm]{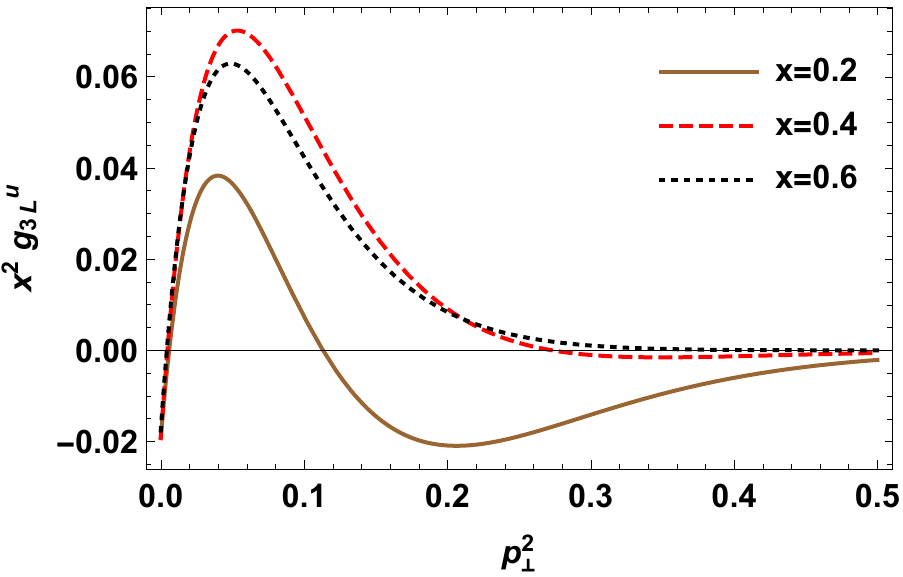}
\hspace{0.05cm}
(b)\includegraphics[width=7.5cm]{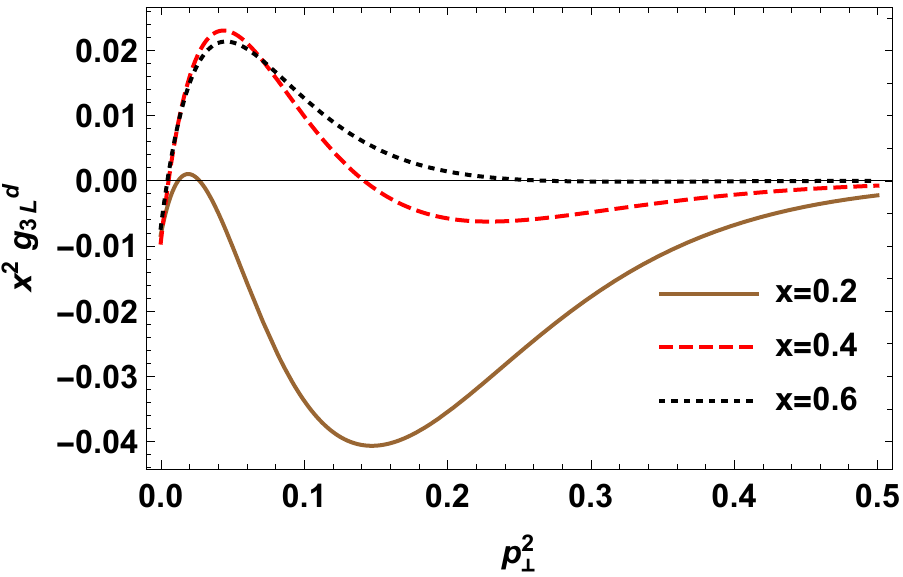}
\hspace{0.05cm}
(c)\includegraphics[width=7.5cm]{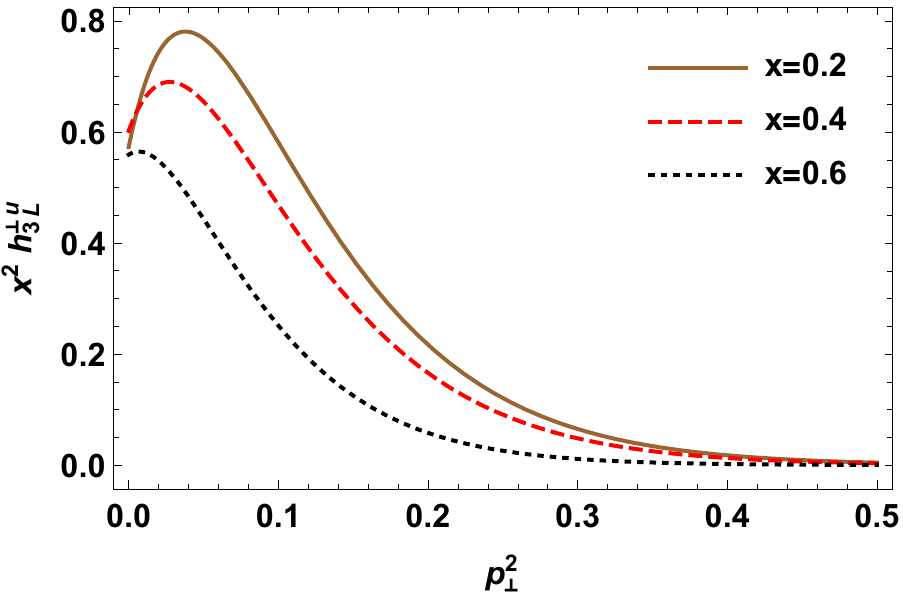}
\hspace{0.05cm}
(d)\includegraphics[width=7.5cm]{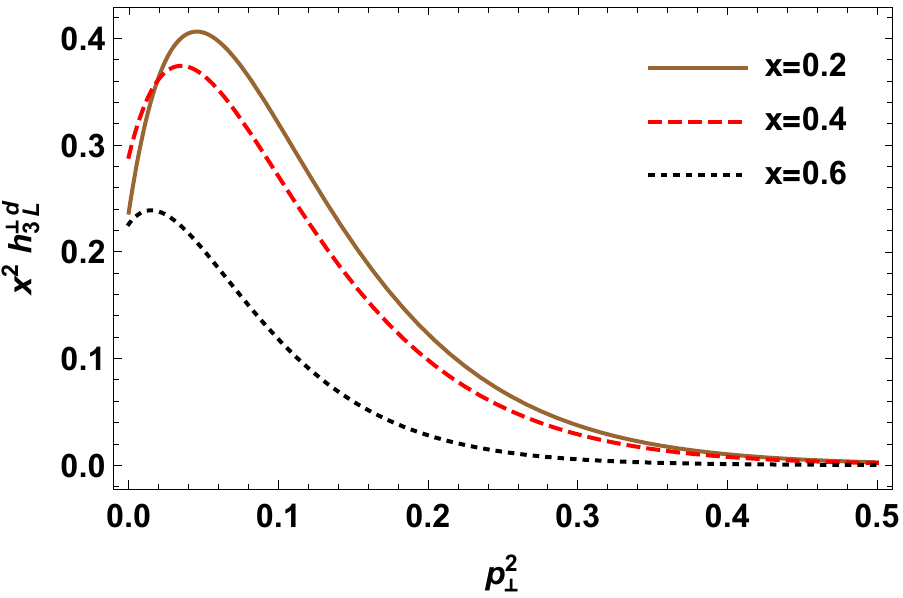}
\hspace{0.05cm}
\end{minipage}
\caption{\label{fig2dvp2} (Color online) The longitudinally polarized TMDs $x^2 g_{3L}^{\nu}(x, {\bf p_\perp^2})$ and $~x^2 h_{3L}^{\perp\nu}(x, {\bf p_\perp^2})$ plotted with respect to ${\bf p_\perp^2}$  at different values of $x$, i.e., $x=0.2$ (brown curve), $x=0.4$ (dashed red curve) and $x=0.6$ (dotted black curve). The left and right column correspond to $u$ and $d$ quarks sequentially.}
\end{figure*}
\begin{figure*}
\centering
\begin{minipage}[c]{0.98\textwidth}
(a)\includegraphics[width=7.5cm]{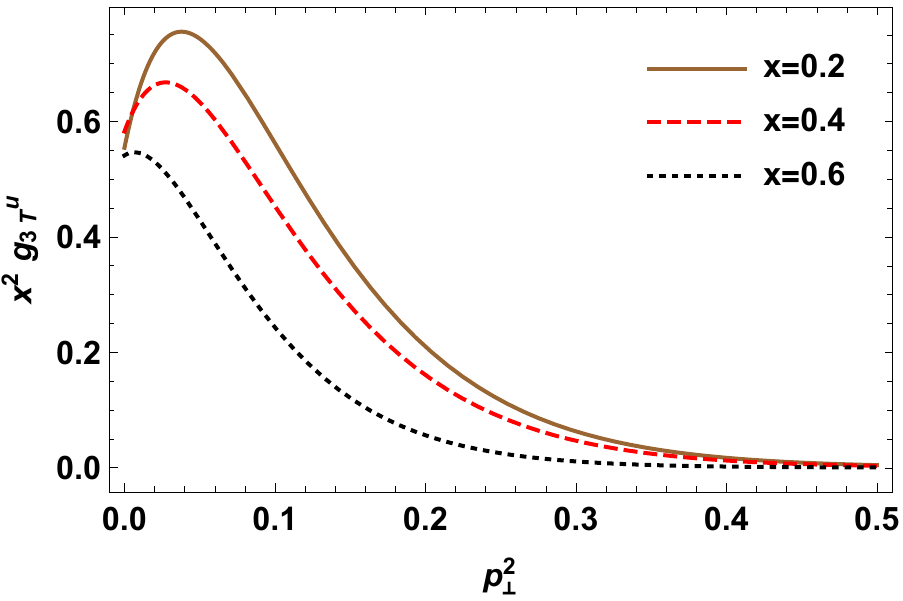}
\hspace{0.05cm}
(b)\includegraphics[width=7.5cm]{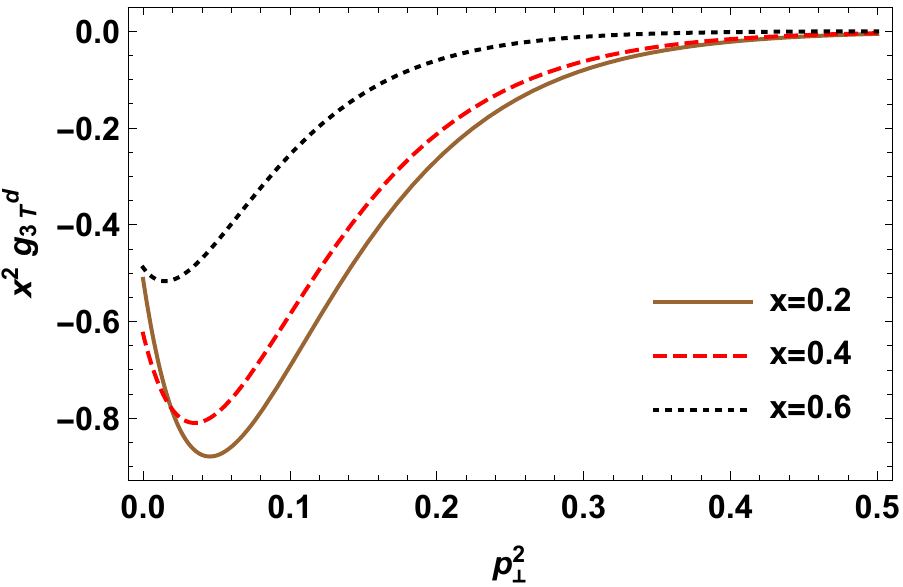}
\hspace{0.05cm}
(c)\includegraphics[width=7.5cm]{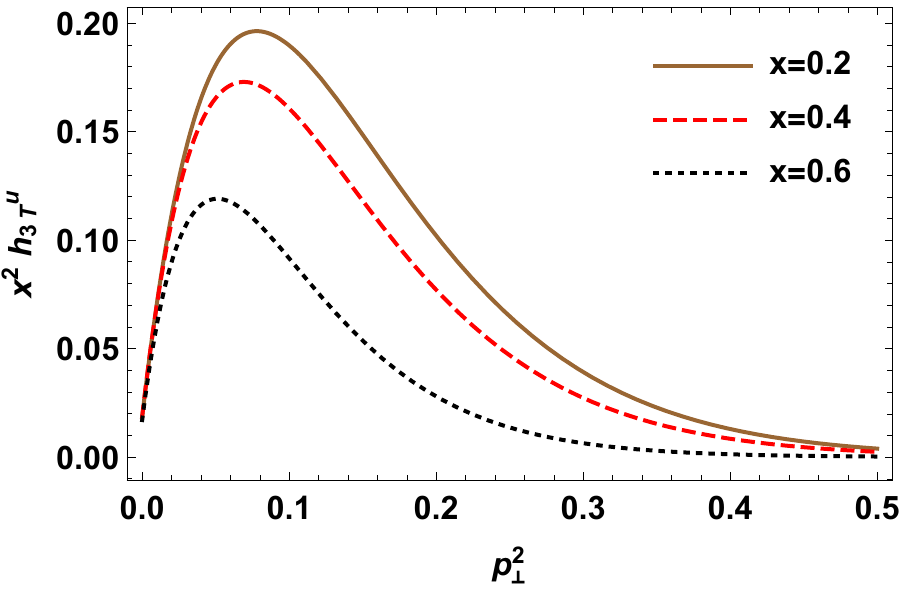}
\hspace{0.05cm}
(d)\includegraphics[width=7.5cm]{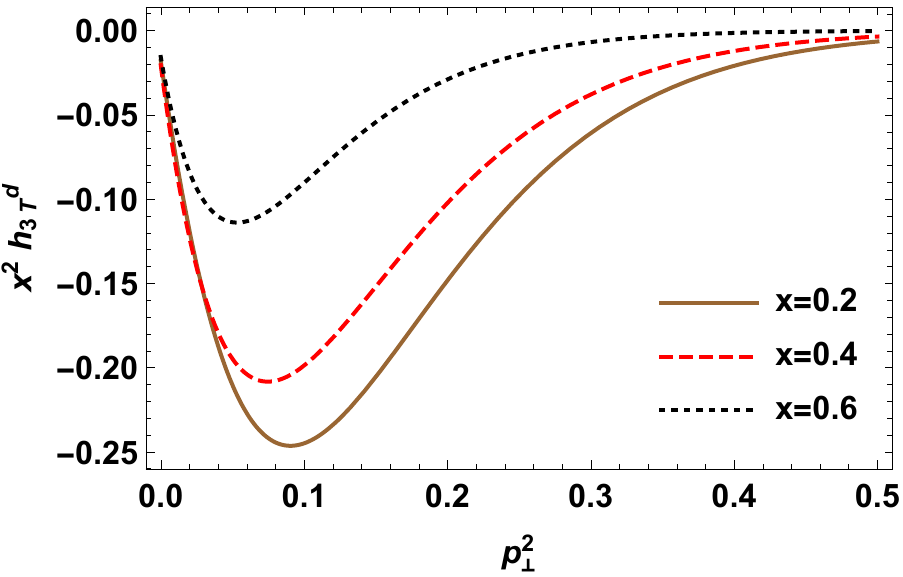}
\hspace{0.05cm}
(e)\includegraphics[width=7.5cm]{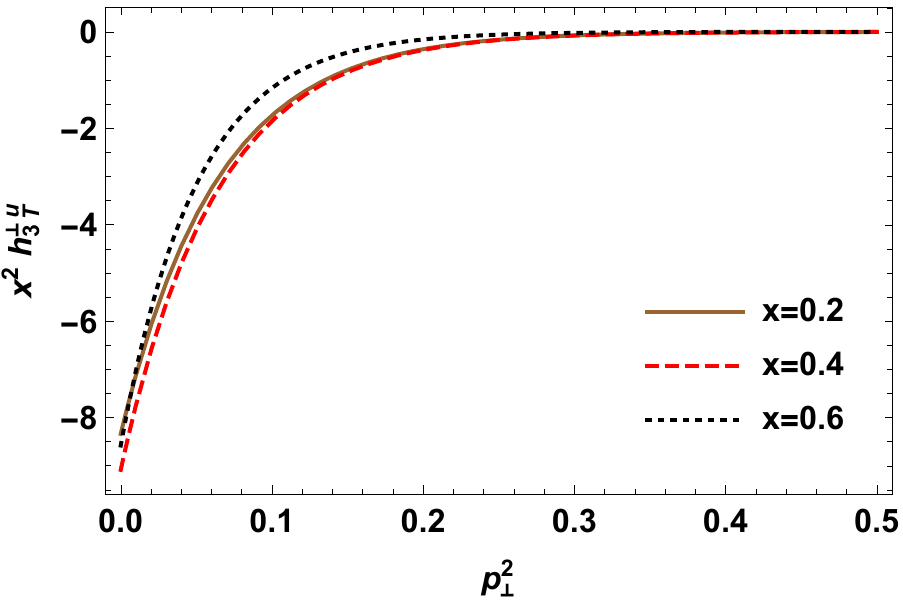}
\hspace{0.05cm}
(f)\includegraphics[width=7.5cm]{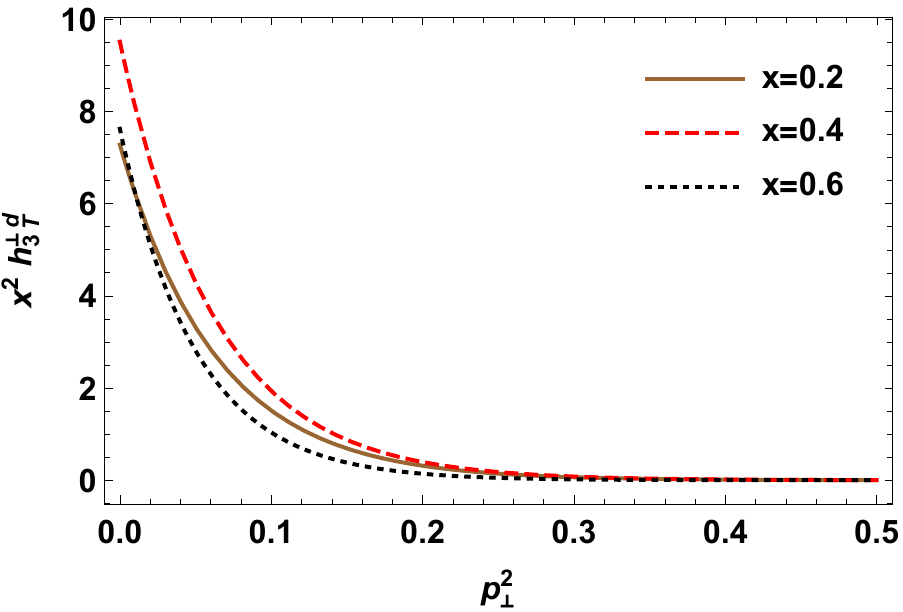}
\hspace{0.05cm}
\end{minipage}
\caption{\label{fig2dvp3} (Color online) The transversely polarized TMDs $x^2 {g}^{\nu  }_{3T}(x, {\bf p_\perp^2}),~x^2 {h}^{\nu  }_{3T}(x, {\bf p_\perp^2})$ and $x^2{h}^{\nu\perp}_{3T}(x, {\bf p_\perp^2})$ plotted with respect to ${\bf p_\perp^2}$  at different values of $x$, i.e., $x=0.2$ (brown curve), $x=0.4$ (dashed red curve) and $x=0.6$ (dotted black curve). The left and right column correspond to $u$ and $d$ quarks sequentially.}
\end{figure*}
\par
To grasp the nature of transverse momentum dependency on twist-4 T-even TMDs, it is necessary to plot them against ${\bf p_\perp^2}$  alone. In Fig. (\ref{fig2dvp1}), ({\ref{fig2dvp2}}) and ({\ref{fig2dvp3}}) the unpolarized \big($x^2 {f}^{\nu  }_3(x, {\bf p_\perp^2})$\big), longitudinally polarized \big($x^2 {g}^{\nu  }_{3L}(x, {\bf p_\perp^2})$ and $x^2 {h}^{\nu\perp}_{3L}(x, {\bf p_\perp^2})$\big) and the transversely polarized \big( $x^2 {g}^{\nu  }_{3T}(x, {\bf p_\perp^2}),$ $x^2 {h}^{\nu  }_{3T}(x, {\bf p_\perp^2})$ and $x^2 {h}^{\nu\perp}_{3T}(x, {\bf p_\perp^2})$\big) TMDs are plotted with respect to ${p_\perp^2}$ for different fixed values of $x$ with different representations, i.e., $x=0.2$ (brown curve), $x=0.4$ (dashed red curve) and $x=0.6$ (dotted black curve). The columns on the left and right correspond to $u$ and $d$ quarks, correspondingly. In Fig. \ref{fig2dvp1} (a) and \ref{fig2dvp1} (b), with the increase in ${p_\perp^2}$ the TMD $x^2 {f}^{\nu  }_3(x, {\bf p_\perp^2})$ first increases and then decreases to show a maxima for both $u$ and $d$ quarks. The longitudinally polarized TMD $x^2 {g}^{\nu  }_{3L}(x, {\bf p_\perp^2})$, with the increase in ${p_\perp^2}$ shows trend as shown in Fig. \ref{fig2dvp2} (a) and \ref{fig2dvp2} (b). With the rise in ${p_\perp^2}$, for both $u$ and $d$ quarks the longitudinally polarized TMD $x^2 {h}^{\nu\perp}_{3L}(x, {\bf p_\perp^2})$, first increases and then decreases to meet the horizontal axis as shown in Fig. \ref{fig2dvp2} (c) and \ref{fig2dvp2} (d). We have plotted the transversely polarized TMD $x^2 {g}^{\nu  }_{3T}(x, {\bf p_\perp^2})$ in Fig. \ref{fig2dvp3} (a) and \ref{fig2dvp3} (b). With the rise in ${\bf p_\perp^2}$, for $u$ ($d$) quarks the TMD first increases (decreases) and then decreases (increases) to meet the horizontal axis. With the increase in ${\bf p_\perp^2}$, the transversely polarized TMD $x^2 {h}^{\nu  }_{3T}(x, {\bf p_\perp^2})$, for $u$ ($d$) quarks first increases (decreases) and then decreases (increases) to meet the ${\bf p_\perp^2}$ axis as shown in Fig \ref{fig2dvp3} (c) \big(\ref{fig2dvp3} (d)\big). In Fig. \ref{fig2dvp3} (e) and \ref{fig2dvp3} (f), the transversely polarized TMD $x^2 {h}^{\nu\perp}_{3T}(x, {\bf p_\perp^2})$ is plotted w.r.t. ${\bf p_\perp^2}$. With the increase in ${\bf p_\perp^2}$, for $u$ ($d$) quarks the amplitude of TMD decreases (increases) to meet the horizontal axis.
\subsection{Integrated TMDs}\label{secitmd}
\begin{figure*}
\centering
\begin{minipage}[c]{0.98\textwidth}
\includegraphics[width=7.5cm]{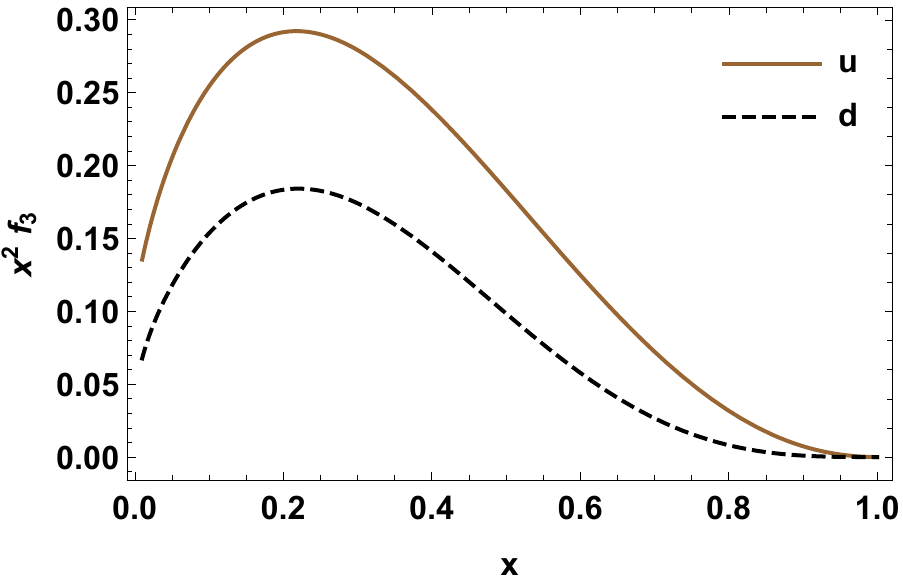}
\hspace{0.05cm}
\end{minipage}
\caption{\label{figitmd1} (Color online) The TMDPDF $x^2 f_{3}^{\nu}(x)$ plotted with respect to $x$. Brown and dashed black curve correspond to $u$ and $d$ quarks sequentially.}
\end{figure*}

\begin{figure*}
\centering
\begin{minipage}[c]{0.98\textwidth}
(a)\includegraphics[width=7.5cm]{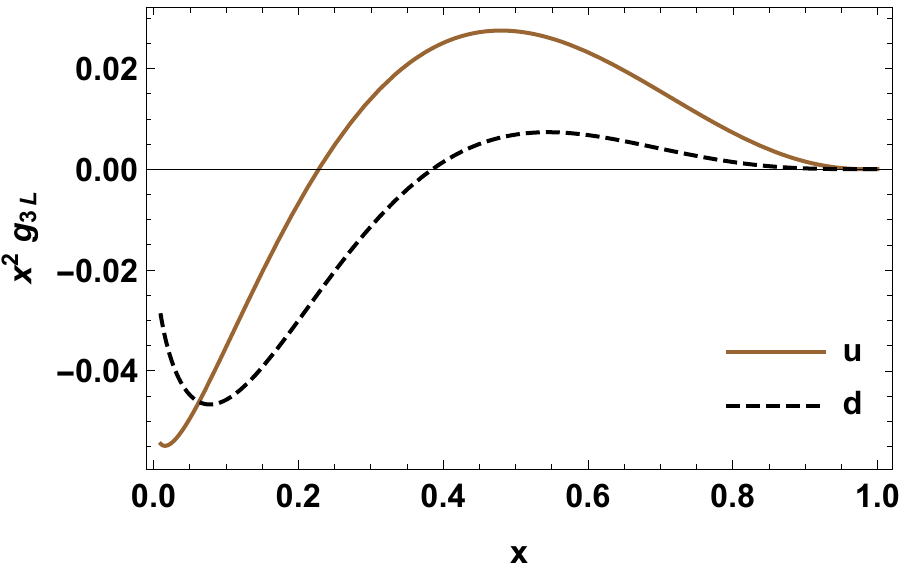}
\hspace{0.05cm}
(b)\includegraphics[width=7.5cm]{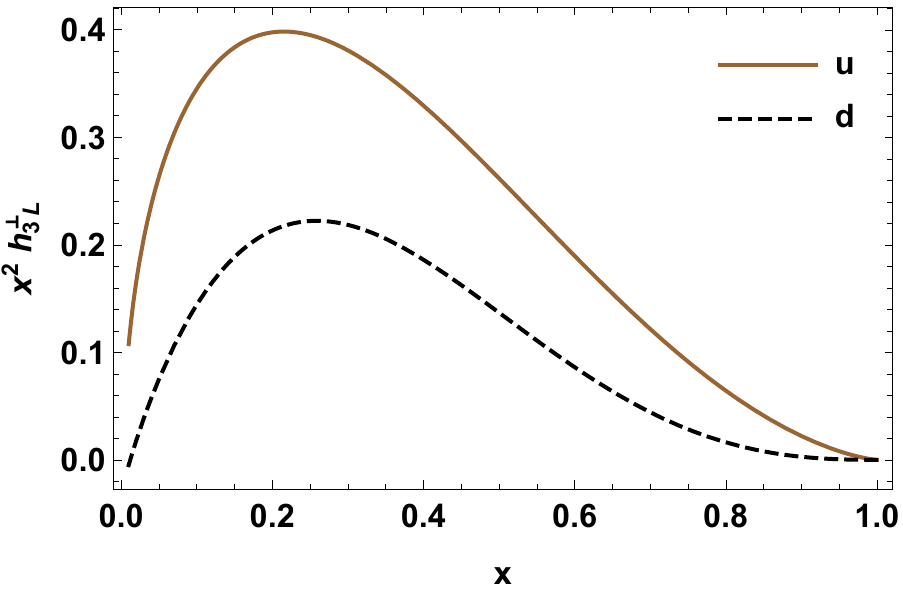}
\hspace{0.05cm}
\end{minipage}
\caption{\label{figitmd2} (Color online) The TMDPDFs $x^2 g_{3L}^{\nu}(x)$ and $~x^2 h_{3L}^{\perp\nu}(x)$ plotted with respect to $x$. Brown and dashed black curve correspond to $u$ and $d$ quarks sequentially.}
\end{figure*}

\begin{figure*}
\centering
\begin{minipage}[c]{0.98\textwidth}
(a)\includegraphics[width=7.5cm]{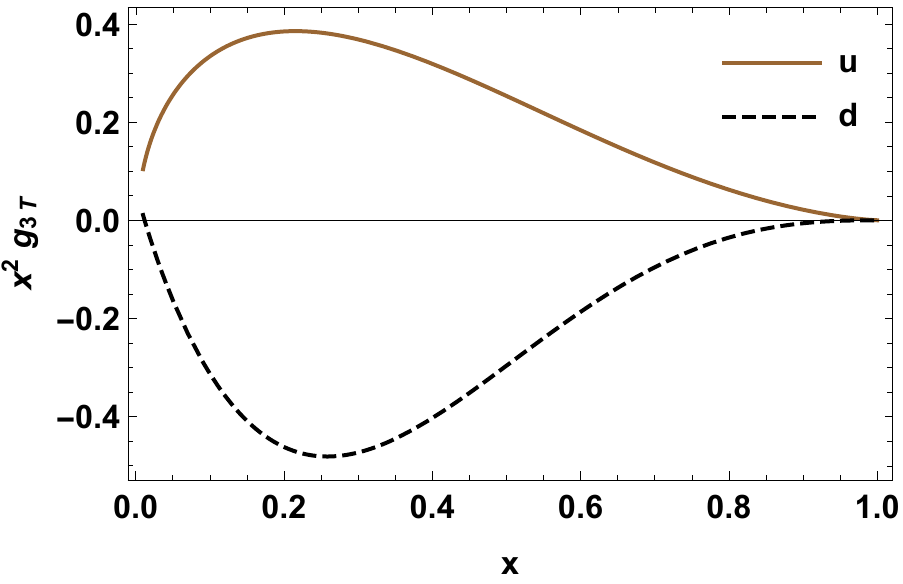}
\hspace{0.05cm}
(b)\includegraphics[width=7.5cm]{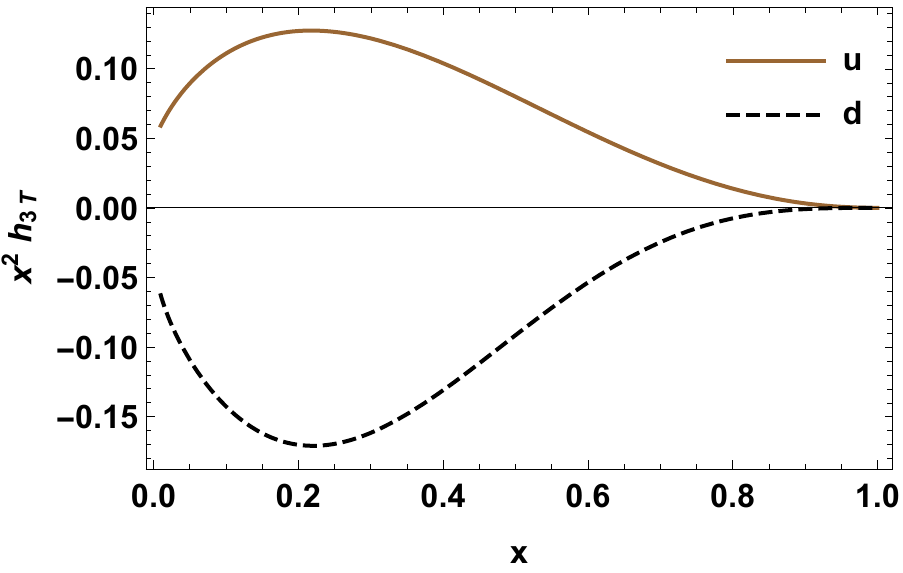}
\hspace{0.05cm}
(c)\includegraphics[width=7.5cm]{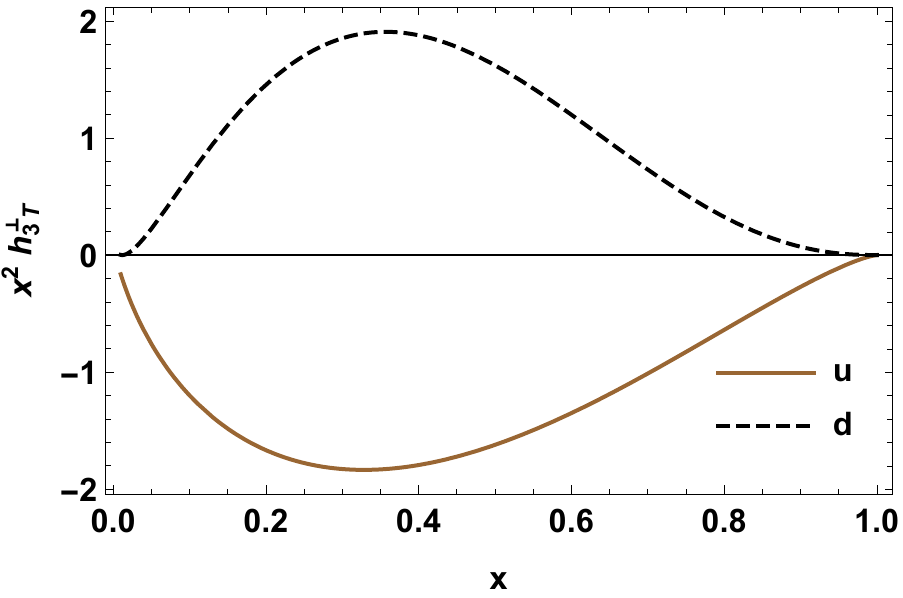}
\hspace{0.05cm}
\end{minipage}
\caption{\label{figitmd3} (Color online) The TMDPDFs $x^2 {g}^{\nu  }_{3T}(x),~x^2 {h}^{\nu  }_{3T}(x)$ and $x^2{h}^{\nu\perp}_{3T}(x)$ plotted with respect to $x$. Golden and dashed black curve corresponds to $u$ and $d$ quarks sequentially.}
\end{figure*}

To gain a clearer image of the proton, the TMDPDFs are created by integrating the TMDs across the transverse momentum of the quark ${\bf p_\perp}$. We have studied their variation with respect to the longitudinal momentum fraction $x$. In Fig. (\ref{figitmd1}), (\ref{figitmd2}) and (\ref{figitmd3}) the unpolarized \big($x^2~{f}^{\nu  }_3(x)$\big), longitudinally polarized \big($x^2~{g}^{\nu  }_{3L}(x)$ and $x^2~{h}^{\nu\perp}_{3L}(x)$\big) and the transversely polarized \big( ${g}^{\nu  }_{3T}(x),~{h}^{\nu  }_{3T}(x)$ and ${h}^{\nu\perp}_{3T}(x)$\big) TMDPDFs are plotted with respect to $x$. Brown and dashed black curve corresponds to $u$ and $d$ quarks sequentially. For unpolarized and longitudinally polarized TMDPDFs, no flip in the sign takes place, but for transversely polarized TMDPDFs a flip in sign is observed in the shape of plot while changing the flavor from $u$ to $d$ quarks or vice versa. In Fig. \ref{figitmd1}, it is spotted that for $u$ ($d$) quarks, with an increase in the longitudinal momentum fraction $x$,  TMDPDF $x^2 f_{3}^{\nu}(x)$ first increases giving a maxima at $x=0.22$ and decreases thereafter. For $u$ quarks, with the increase in the longitudinal momentum fraction $x$, the longitudinally polarized TMDPDF $x^2 g_{3L}^{\nu}(x)$ rises to the positive region and then decreases to meet $x$ axis by showing maxima at $x=0.48$ as shown in Fig. \ref{figitmd2} (a).
In the same figure for $d$ quarks, with a rise in $x$, the TMDPDF decreases and then increases to reveal the minima at $x=0.08$ and at the end again decreases to show its maxima at $x=0.55$. In Fig. \ref{figitmd2} (b), it is observed that for $u$ ($d$) quarks, with an increase in the longitudinal momentum fraction $x$,  TMDPDF $x^2 h_{3L}^{\perp\nu}(x)$ first increases giving a maxima at $x=0.22~(0.26)$ and decreases thereafter. In Fig. \ref{figitmd3} (a), it is observed that for $u$ ($d$) quarks, with an increase in the longitudinal momentum fraction $x$,  TMDPDF $x^2 {g}^{\nu  }_{3T}(x)$ first increases giving a maxima (minima) at $x=0.22~(0.26)$ and decreases (increases) after that. With an increase in the longitudinal momentum fraction $x$,  TMDPDF $x^2 {h}^{\nu  }_{3T}(x)$ first increases giving a maxima (minima) at $x=0.22~(0.22)$ and decreases (increases) after that for $u$ ($d$) quarks as shown in Fig. \ref{figitmd3} (b). The TMDPDF $x^2 {h}^{\nu\perp}_{3T}(x)$ first decreases (increases) and then increases (decreases) with increase in $x$ and giving minima (maxima) at $x=0.33~(0.36)$ for $u$ ($d$) quarks as shown in  Fig. \ref{figitmd3} (c). Moreover, close inspection of TMDPDFs reveal that their trend is similar to that of respective TMDs.

\section{Conclusion}\label{seccon}
In this work, we have presented the study of unpolarized $\bigg(f_{3}^{\nu}(x, {\bf p_\perp^2}) \bigg)$,
 longitudinally polarized $\bigg( g_{3L}^{\nu}(x, {\bf p_\perp^2})$ and $ h_{3L}^{\perp\nu}(x, {\bf p_\perp^2})\bigg)$ and transversely polarized \bigg( ${g}^{\nu  }_{3T}(x, {\bf p_\perp^2}),~{h}^{\nu  }_{3T}(x, {\bf p_\perp^2})$ and 
${h}^{\nu\perp}_{3T}(x, {\bf p_\perp^2})$\bigg) twist-4 T-even TMDs, by systematic calculations in a framework of LFQDM. After acquiring the overlap form of TMDs by exploiting the unintegrated SIDIS quark-quark correlator, we have presented the explicit expressions for both the cases of diquark being a scalar or a vector. We have shown the 2-D and 3-D variation of these TMDs with longitudinal momentum fraction $x$ and transverse momentum ${\bf p_\perp^2}$ for both $u$ and $d$ quarks. For unpolarized and longitudinally polarized TMDs, no flip in the sign takes place in the shape of plot while changing the flavor from $u$ to $d$ quarks or vice versa, whereas for transversely polarized TMDs a flip in sign is observed. At large value of ${\bf p_\perp^2}$, we have found the amplitude of TMDs to be very low, and in fact it is significantly negligible when ${\bf p_\perp^2}$ is above  $0.3~\mathrm{GeV}^2$. The reason lies in the model itself, because the wave functions using which we have expressed our results are exponential of negative ${\bf p_\perp^2}$ \big(i.e., $\exp\big[-\delta^\nu\frac{\bfp^2}{2\kappa^2}\frac{\log(1/x)}{(1-x)^2}\big]$\big), with the dependence on other parameters as well. But, when the value of ${\bf p_\perp^2}$ is greater than and equal to $0.3~\mathrm{GeV}^2$, the dominance of exponentially decreasing factor is observed.  The TMD amplitude of $u$ quarks is always greater than that of $d$ quarks. As the selected value of ${\bf p_\perp^2}$ is increased in the plot of TMD versus longitudinal momentum fraction, the amplitude of TMD drops. We have obtained the TMDPDFs by integrating the TMDs over the transverse momentum of quark ${\bf p_\perp}$. We have studied their variation with respect to the longitudinal momentum fraction $x$. Sign of plot for unpolarized and longitudinally polarised TMDPDFs is the same for both flavors, however the sign of transversely polarised TMDPDFs flips when $u$ quarks are replaced by $d$ quarks. Additionally, it has been noted that the trend of TMDPDF is analogous to that of their corresponding TMD.
\par
We have also provided the model relations of twist-4 T-even TMDs with the leading twist T-even TMDs. To be specific, we have expressed our unpolarized twist-4 T-even TMD $f_{3}^{\nu}(x, {\bf p_\perp^2})$ in the form of unpolarized leading twist T-even TMD ${f}^{\nu  }_1(x, {\bf p_\perp^2})$ and  expressions of our twist-4 T-even longitudinally polarized TMDs $\bigg(g_{3L}^{\nu}(x, {\bf p_\perp^2})$ and $ h_{3L}^{\perp\nu}(x, {\bf p_\perp^2})\bigg)$ in the form of leading twist T-even longitudinally polarized TMDs  $\bigg({h}^{\nu\perp}_{1L}(x, {\bf p_\perp^2})$ and ${g}^{\nu  }_{1L}(x, {\bf p_\perp^2})\bigg)$. The relations of transversely polarized twist-4 T-even TMDs \bigg(${g}^{\nu  }_{3T}(x, {\bf p_\perp^2}),~{h}^{\nu  }_{3T}(x, {\bf p_\perp^2})$ and ${h}^{\nu\perp}_{3T}(x, {\bf p_\perp^2})$\bigg) with transversely polarized leading twist T-even TMDs \bigg(${g}^{\nu  }_{1T}(x, {\bf p_\perp^2}),~{h}^{\nu}_{1T}(x, {\bf p_\perp^2}),~{h}^{\nu\perp}_{1T}(x, {\bf p_\perp^2}) $ and ${h}^{\nu}_{1}(x, {\bf p_\perp^2})$\bigg) have been provided. These model oriented relations might become one of the most captivating results of LFQDM. The relation of unpolarized TMD has also been obtained in previous studies and it is in sync with our result. Future studies will throw more light on the model independent relations of these TMDs.\par

We have tabulated the results of average transverse momenta and average transverse momenta square for our twist 4 T-even TMDs and compared them to the results from our model's leading twist T-even TMDs. All results are in units of the respective value for $f_1^{\nu}$, which is $ \langle p_\perp \rangle^{u} =0.23~\mathrm{GeV}$, $\langle p_\perp \rangle^{d}=0.24~\mathrm{GeV}, \langle p_\perp^2 \rangle^{u} =0.066~\mathrm{GeV}^2$, $\langle p_\perp^2 \rangle^{d}=0.075~\mathrm{GeV}^2$ for reference. Their detailed inspection revealed that, just as leading twist TMD $f_{1}^{\nu}(x, {\bf p_\perp^2})$  and $ h_{1T}^{\nu}(x, {\bf p_\perp^2}) $ have the same average transverse momentum and average square transverse momentum value, their parallel twist-4 partners $ f_{3}^{\nu}(x, {\bf p_\perp^2}) $ and $ h_{3T}^{\nu}(x, {\bf p_\perp^2}) $ demonstrate a similar pattern.
Similarly, twist-4 TMDs $g_{3T}^{\nu}(x, {\bf p_\perp^2})$ and $h_{3L}^{\perp\nu}(x, {\bf p_\perp^2})$ have equal values just like the behaviour between their leading twist companions $g_{1T}^{\nu}(x, {\bf p_\perp^2})$ and $h_{1L}^{\perp\nu}(x, {\bf p_\perp^2})$. Along with it, the values of average transverse momentum and average square transverse momentum for TMD ${f}^{\nu  }_3(x, {\bf p_\perp^2})$ from LFCQM have been compared with our results. The values obtained in our model are flavor dependent, whereas in LFCQM they are flavor independent. Values of $ \langle p_\perp \rangle^{u}$ and $ \langle p_\perp \rangle^{d}$ are in sync but $\langle p_\perp^2 \rangle^{u}$ and $\langle p_\perp^2 \rangle^{d}$ values in our model are slightly smaller but comparable with the LFCQM results.
\par In conclusion, the contributions of twist-4 T-even TMDs and associated TMDPDFs are rather considerable and they have been the focus of research in DIS studies such as HERMES and those conducted in J-Lab. 
Future applications of the light-front method to calculate higher-twist sea quark and gluon TMDs would be fascinating. We feel that defining and employing models capable of capturing, beyond the Gaussian approximation, the diverse combinations of parton and nucleon polarizations is crucial to advancing our methodology for hadron 3D imaging. This is attributed to the reason that first moments or higher-twist TMD densities are directly incorporated into the definition of the collinear matching of related lower-twist TMD densities. In this regard, one would want the entire set of both quark (valence and sea) and gluon higher-twist TMDs to obtain the collinear input entering the matching term of the lower-twist TMDs. Apart from that, it would be very exciting to do detailed studies on the higher twist generalized transverse momentum dependent distributions and  Wigner distributions.
\section{Acknowledgement}
H.D. would like to thank the Science and Engineering Research Board, Department of Science and Technology, Government of India through the grant (Ref No. MTR/2019/000003) under MATRICS scheme for financial support.



%
\end{document}